\documentclass[12pt,a4paper,english]{article}
\usepackage[T1]{fontenc}
\usepackage[latin2]{inputenc}
\usepackage[english]{babel}
\usepackage{amsmath}
\usepackage{amsfonts}
\usepackage{indentfirst}
\usepackage[dvips]{graphicx}

\newcommand{\bea}{\begin{eqnarray}}
\newcommand{\eea}{\end{eqnarray}}
\newcommand{\be}{\begin{equation}}
\newcommand{\ee}{\end{equation}}

\newcommand{\Z}{{\mathbb Z}}

\newcommand{\C}{{\mathbb C}}

\newcommand{\cZ}{{\mathcal{Z}}}

\newcommand{\Li}{{\rm Li}}

\def\Tr{{\rm Tr \,}}

\def\G{\Gamma}

\newcommand{\cH}{{\cal H }}

\begin{document}

\sloppy

\begin{flushright}
\begin{tabular}{l}
CALT-68-2835 \\

\\ [.3in]
\end{tabular}
\end{flushright}

\begin{center}
\Large{ \bf BPS states, crystals and matrices}
\end{center}

\begin{center}

\bigskip

\bigskip 

Piotr Su{\l}kowski

\bigskip

\bigskip

\medskip 

\emph{California Institute of Technology, Pasadena, CA 91125, USA} \\ 

\medskip

\emph{and}

\medskip

\emph{Faculty of Physics, University of Warsaw \\ 
 ul. Ho{\.z}a 69, 00-681 Warsaw, Poland}

\bigskip

\emph{psulkows@theory.caltech.edu}

\bigskip

\smallskip
 \vskip .4in \centerline{\bf Abstract}
\smallskip

\end{center}

We review free fermion, melting crystal and matrix model representations of wall-crossing phenomena on local, toric Calabi-Yau manifolds. We consider both unrefined and refined BPS counting of closed BPS states involving D2 and D0-branes bound to a D6-brane, as well as open BPS states involving open D2-branes ending on an additional D4-brane. Appropriate limit of these constructions provides, among the others, matrix model representation of refined and unrefined topological string amplitudes.


\newpage 

\tableofcontents

\newpage 


\section{Introduction}

This review is devoted to some aspects of counting of BPS states in a system of D$p$-branes, with even $p$, in type IIA string compactifications. The problems of BPS counting span a vast area of research in supersymmetric gauge and string theories. Their important feature is a special, non-constant character of BPS multiplicities: their values depend on various moduli and jump discontinuously along some special loci in the corresponding moduli space, so called \emph{walls of marginal stability}. The pattern of these jumps follows \emph{wall-crossing formulas}, found from physical perspective by Denef and Moore \cite{DenefMoore} and, in more general context, formulated mathematically by Kontsevich and Soibelman \cite{KS}. The regions of the moduli space in between walls of marginal stability, in which BPS multiplicities are (locally) constant, are called \emph{chambers}.

The BPS states we are interested in, and which we will refer to as \emph{closed BPS states}, arise as bound states of a single D6-brane with arbitrary number of D0 and D2-branes wrapping cycles of a toric Calabi-Yau space. More generally, we will also consider \emph{open BPS states}, which arise when an additional D4-brane spans a lagrangian submanifold inside the Calabi-Yau space and supports open D2-branes attached to it. The closed and open BPS states give rise, respectively, to single-particle states in the effective four-dimensional and two-dimensional theory (in remaining, space-time filling directions of, respectively, D6 and D4-branes). In this context the character of BPS multiplicities can be understood in much detail, and it relates to other interesting exactly solvable models: free fermions, crystal, and matrix models. In brief, these connections arise as follows. Firstly, BPS states we consider turn out to be in one-to-one correspondence with configurations of certain statistical models of melting crystals. The structure of these crystals depends on geometry of the underlying Calabi-Yau space, as well as on the chamber one is considering. In consequence BPS counting functions, upon appropriate identification of parameters, coincide with generating functions of melting crystals. It turns out that the structure of these crystals can be given a free fermion representation. Furthermore, once such free fermion formulation is known, it can also be represented in terms of matrix models. Connection with vast theory of matrix models has many interesting mathematical and physical consequences and allows to shed new light on wall-crossing phenomena. The aim of this review is to explain these connections.

The BPS generating functions which we consider are intimately related to topological string amplitudes on corresponding Calabi-Yau spaces. This relation is most transparent in the physical derivation discussed in section \ref{sec-BPS}, which relies on lifting the D-brane system to M-theory. The M-theory viewpoint makes contact with original formulation of closed topological strings by Gopakumar and Vafa \cite{GV-I,GV-II}, and open topological strings by Ooguri and Vafa \cite{OV99}. In particular, in one specific, so called \emph{non-commutative chamber}, the BPS generating function is given as the modulus square of the topological string partition function. In all other chamber BPS generating functions can be uniquely determined from that non-commutative result. There is also another special, so called \emph{commutative chamber}, in which BPS generating function coincides (up to the factor of MacMahon function) with the topological string partition function. For toric manifolds which we consider, such topological string amplitudes can be constructed, among the other, by means of the powerful topological vertex formalism \cite{topvertex}. Relation to crystal models was in fact first understood in this topological string chamber, starting with the seminal work of Okounkov, Reshetikhin and Vafa \cite{ok-re-va}, and quickly followed by \cite{foam,MNOP-I}. One advantage of the formalism presented in this review is the fact that it allows to construct matrix model representation of all these generating functions (so, in particular, matrix model representation of topological string amplitudes).

In more detail, we will consider generating functions of D2 and D0-branes bound to a single D6-brane of the following form form 
\be 
\cZ_{BPS}(q_s, Q) = \sum_{\alpha, \beta} 
\Omega(\alpha, \beta) q_s^\alpha Q^\beta, 
\label{BPSpartition}
\ee
where $\alpha \in \mathbb{Z}$ is D0-brane charge, and $\beta \in H_2(X, \mathbb{Z})$ is D2-brane charge. Multiplicities $\Omega(\alpha, \beta)$ jump when central charges (which itself are functions of K{\"a}hler moduli) of building blocks of a bound state align, and therefore these generating functions are locally constant functions of K{\"a}hler moduli. Along the walls of marginal stability the degeneracies $\Omega(\alpha,\beta)$ change and indeed obey wall-crossing formulas of \cite{DenefMoore,KS} mentioned above.

If there is an additional D4-brane which spans a lagrangian submanifold inside the Calabi-Yau space, in addition to the above \emph{closed} BPS states, one can consider also \emph{open} BPS states of D2-branes with boundaries ending on a one-cycle $\gamma$ on this D4-brane. In this case the BPS states arise on the remaining two-dimensional world-volume of the D4-brane. The holonomy of the gauge field along $\gamma$ provides another generating parameter $z$, so that open BPS generating functions take form
\be 
\cZ_{BPS}^{open}(q_s, Q) = \sum_{\alpha, \beta,\gamma} 
\Omega(\alpha, \beta) q_s^\alpha Q^\beta z^{\gamma}.
\label{BPSpartition-open}
\ee
As we will show, generating functions of such open BPS states can be identified with integrands of matrix models mentioned above.

One more important aspect of BPS counting is referred to as \emph{refinement}, and amounts to refining BPS counting by introducing one more parameter, customarily denoted $\beta$. The refinement can be introduced from several perspectives which give rise to identical results, however their fundamental common origin is still not fully understood. We will introduce refinement by distinct counting of states with different $SU(2)$ spins inside spacetime $SO(4)$ rotation group in the generating function (\ref{BPSpartition}). In \cite{RefMotQ} it was argued that this physical viewpoint should agree with the mathematical counterpart of motivic deformation \cite{KS}, and also a refined version of a crystal model was constructed. Another notion of refinement arises in Nekrasov partition functions, which are defined in a non-trivial gravitational (so called $\Omega$-) background parametrized by two parameters $\epsilon_1$ and $\epsilon_2$ introduced by Nekrasov \cite{Nek} and further developed by Nekrasov and Okounkov \cite{Nek-Ok}. Nekrasov partition functions can also be defined for five-dimensional gauge theories and then they agree with topological string amplitudes. In particular, the formalism of the topological vertex \cite{topvertex} has also been extended to the refined context in \cite{refined-vertex}, and shown to reproduce relevant Nekrasov partition functions. Also BPS generating functions, in the limit of commutative chamber, are known to reproduce refined topological string amplitudes with $\beta=-\epsilon_1/\epsilon_2$ \cite{Nagao-open}. However the worldsheet definition of refined topological string amplitudes is not fully understood.   

As an exemplary and, hopefully, inspiring application of the entire formalism presented in this review, in the final section \ref{sec-refine} we derive matrix model representation of the refined topological string partition function for the conifold. 
The refined matrix model which we find has a standard measure, however its potential is deformed by $\beta$-dependent terms. It is obtained by constructing appropriate refined crystal model and free fermion representation, and subsequently reformulating this representation in matrix model form. Finally, taking the limit of the commutative chamber, we obtain matrix model representation of the refined topological string amplitude. Even though we demonstrate this result in the conifold case, with some technical effort it can be generalized to other toric manifolds which we consider.\footnote{As we recall in section \ref{sec-refine}, refined topological string amplitudes were also postulated to be reproduced by another type of matrix models, so-called $\beta$-deformed ones (whose Vandermonde measure is deformed by raising it to power $\beta$); however explicit computations showed that this cannot be the correct representation of refined amplitudes.}


\subsection*{Short literature guide}

The literature on the topics presented in this paper is extensive and still growing, and we unavoidably mention just a fraction of important developments. The relation between Donaldson-Thomas invariants for the non-commutative chamber of the conifold was first found by Szendroi \cite{Szendroi}. It was generalized to orbifolds of $\mathbb{C}^3$, and related to free fermion formalism, by Bryan and Young \cite{YoungBryan}. The relation to free fermions and crystals was extended to a large class of toric manifolds without compact four-cycles \cite{ps-pyramid,NagaoVO}. These developments were accompanied by other mathematical works \cite{NagaoNakajima,Nagao-strip}. 

In parallel to the above mentioned mathematical activity, wall-crossing phenomena for local Calabi-Yau manifolds were analyzed from physical viewpoint. The analysis of non-trivial BPS counting for the conifold was described by Jafferis and Moore in \cite{JaffMoore}. This and more general cases were related to quivers and crystal models in \cite{ChuangJaff,OY1}. Derivation of BPS degeneracies from M-theory viewpoint and relation to closed topological strings were discussed in \cite{WallM}, and generalized to open BPS counting in \cite{AY-owc,DSV,NagaoYamazaki,openWalls}. Relations to matrix models, discussed for plane partitions with some other motivation in \cite{EynardTASEP}, were extended to other crystal models relevant for BPS counting in \cite{OSY}, and also in \cite{SzaboTierz-DT}. Subsequently it was related to open BPS counting in \cite{openWalls}. Refined BPS counting was related to crystal models in \cite{RefMotQ,Nagao-open}, and corresponding matrix models were constructed in \cite{refine}.

Let us also mention some other, related works devoted to crystals, random matrices and free fermions. Many new results and nice summaries concerning these topics are given in works by Okounkov \cite{ok1,ok2,ok3} and Okounkov and Reshetikhin \cite{ok-res,ok-res-2}. The fermionic construction of MacMahon function for $\C^3$ and the topological vertex was originally presented in the foundational work of Okounkov, Reshetikhin and Vafa \cite{ok-re-va}, and its relation to open topological strings and more complicated Calabi-Yau manifolds were discussed in \cite{va-sa,ps,bubblingCY}. Newer ideas, analyzing more complicated systems involving D4-branes, were presented in \cite{QuiversCrystals,D4D2D0}. More expository presentations of various aspects described here can be found in \cite{ps-phd,MY-phd}. A general introduction to mathematical and physical aspects of mirror symmetry can be found in \cite{MirrorSymmetry}.

\subsection*{Plan}

The plan of this paper is as follows.    
In section \ref{sec-BPS} we introduce BPS generating functions and present one possible derivation of their form, which relies on the M-theory interpretation of a D-brane system, following \cite{WallM,AY-owc,DSV,openWalls}. In section \ref{sec-background} we provide a little mathematical background and introduce notation pertaining to toric Calabi-Yau manifolds, free fermion formalism, and matrix models. In section \ref{sec-BPSfermions} we introduce fermionic formalism for BPS generating functions and present corresponding crystal models, building on earlier ideas of \cite{ok-re-va,YoungBryan} and following \cite{ps-pyramid}. In section \ref{sec-matrix} we reformulate the problem of closed BPS counting in terms of matrix models and relate it to open BPS counting \cite{openWalls,OSY}. In section \ref{sec-refine} we refine our analysis, present refined BPS generating functions and crystals \cite{RefMotQ}, and construct corresponding refined matrix models \cite{refine}. 


\section{BPS generating functions}  \label{sec-BPS}

In this section we introduce generating functions of BPS states of D-branes in toric Calabi-Yau manifolds. Our task in the rest of this paper is to provide interpretation of these generating functions in terms of free fermions, melting crystals and matrix models. These generating functions can be derived using wall-crossing formulas, as was done first in the unrefined \cite{JaffMoore} and refined \cite{RefMotQ} conifold case, and later generalized to arbitrary geometry without compact four-cycles in \cite{NagaoNakajima,Nagao-strip}. On the other hand, we will focus on a simpler physical derivation of BPS generating functions which uses the lift of the D-brane system to M-theory \cite{WallM}. This also makes contact with M-theory interpretation of topological string theory, and allows to express BPS counting functions in terms of topological string amplitudes. Moreover this M-theory derivation can be extended to the counting of open BPS states, i.e. open D2-branes attached to additional D4-brane, which we are also interested in \cite{AY-owc,DSV,openWalls}. 

We start this section by reviewing the M-theory derivation of (unrefined) closed and open BPS generating functions. Then, to get acquainted with a crystal interpretation of these generating functions, we discuss their crystal interpretation in simple cases of $\C^3$ and conifold. Later, using fermionic interpretation, we will generalize this crystal representation to a large class of toric manifolds without compact four-cycles.


\subsection{M-theory derivation}   \label{ssec-M}

We start by considering a system of D2 and D0-branes bound to a single D6-brane in type IIA string theory. It can be reinterpreted in M-theory as follows \cite{WallM}. When additional $S^1$ is introduced as the eleventh dimension transversely to the D6-brane, then this D6-brane transforms into a geometric background of a Taub-NUT space with unit charge \cite{DVV-Mduality}. 
The Taub-NUT space is a circle fibration over $\mathbb{R}^3$, with a circle $S^1_{TN}$ attaining a fixed radius $R$ at infinity, and shrinking to a point in the location of the original D6-brane. From M-theory perspective bound states involving D2 and D0-branes are interpreted as M2-branes with momentum on a circle. Therefore the counting of original bound states to the D6-brane is reinterpreted as the counting of BPS states of M2-branes in the Taub-NUT space. While in general this is still a nontrivial problem, for the purpose of counting BPS degeneracies we can take advantage of their invariance under continuous deformations of the Taub-NUT space, in particular under deformations of the radius $R$. We can therefore consider taking this radius to infinity, whereupon BPS counting is reinterpreted in terms of a gas of particles in $\mathbb{R}^5$. To make the problem fully tractable we have to ensure that the particles are non-interacting, which would be the case if moduli of the Calabi-Yau would be tuned so that M2-branes wrapped in various ways would have aligned central charges. This can be achieved when K{\"a}hler parameters of the Calabi-Yau space are tuned to zero. However, to avoid generation of massless states, at the same time one has to include non-trivial fluxes of the M-theory three-form field through the two-cycles of the Calabi-Yau and $S^1_{TN}$. In type IIA this results in the $B$-field flux $B$ through two-cycles of Calabi-Yau. Finally, to avoid creation of the string states arising from M5-branes wrapping four-cycles in Calabi-Yau, we simply restrict considerations to manifolds without compact four-cycles.  For a state arising from D2-brane wrapping a class $\beta$ the central charge then reads
\be
Z(l,\beta) = \frac{1}{R}(l + B\cdot \beta),         \label{Zcentral}
\ee
where $l$ counts the D0-brane charge, which is taken positive to preserve the same supersymmetry. 

Under the above conditions, the counting of D6-D2-D0 bound states is reinterpreted in terms of a gas of particles arising from M2-branes wrapped on cycles $\beta$. The excitations of these particles in $\mathbb{R}^4$, parametrized by two complex variables $z_1,z_2$, are accounted for by the modes of the holomorphic field 
\be
\Phi(z_1,z_2) = \sum_{l_1,l_2} \alpha_{l_1,l_2} z_1^{l_1} z_2^{l_2}.    \label{phi-z1z2}
\ee
Decomposing the isometry group of $\mathbb{R}^4$ as $SO(4)=SU(2)\times SU(2)'$ there are $N_{\beta}^{m,m'}$ five-dimensional BPS states of intrinsic spin $(m,m')$. We are interested in their net number arising from tracing over $SU(2)'$ spins
$$
N_{\beta}^m = \sum_{m'} (-1)^{m'} N_{\beta}^{m,m'}.
$$
The total angular momentum of a given state contributing to the index is $l=l_1+l_2+m$. Finally, in a chamber specified by the moduli $R$ and $B$, the invariant degeneracies can be expressed as the trace over the corresponding Fock space
\bea
\cZ_{BPS} & = & \Big( \textrm{Tr}_{Fock} q_s^{Q_0} Q^{Q_2} \Big)\, |_{chamber} = \nonumber \\
& = & \prod_{\beta,m} \prod_{l_1+l_2 = l} (1-q_s^{l_1+l_2+m} Q^{\beta})^{N_{\beta}^{m}} \, | _{chamber} \nonumber \\
& = & \prod_{\beta,m} \prod_{l=1}^{\infty} (1-q_s^{l+m} Q^{\beta})^{l N_{\beta}^{m}} \, | _{chamber},     \label{cZ-chamber}
\eea
where the subscript $chamber$ denotes restriction to those factors in the above product, which represent states which are mutually BPS
\be
Z(l,\beta)  > 0 \qquad \qquad \Leftrightarrow \qquad \qquad q_s^{l+m} Q^{\beta} < 1.       \label{Zpositive}
\ee
As usual, $Q=e^{-T}$ and $q_s=e^{-g_s}$ above encode respectively the K{\"a}hler class $T$ and the string coupling $g_s$ (we wish to distinguish carefully $q_s$ which encodes string coupling, from a counting parameter $q$ which will arise in what follows in crystal interpretation).
%
The above condition on central charges is crucial in determining a  particular form of the BPS generating functions. If we would restrict products in the formula (\ref{cZ-chamber}) to factors with only positive $\beta$, we would get (up to possibly some factor of MacMahon function) the Gopakumar-Vafa representation of the topological string amplitude. With all negative and positive values of $\beta$ we would get modulus square of the topological string partition function. Therefore the upshot of \cite{WallM} is that in general the above BPS generating function can be expressed in terms of the closed topological string partition function
\be
\cZ_{BPS} = \cZ_{top}(Q) \cZ_{top}(Q^{-1}) |_{chamber},     \label{ZBPS-Ztop}
\ee
where chamber restriction is to be understood as picking up only those factors in Gopakumar-Vafa product representation of $\cZ_{top}$ for which (\ref{Zpositive}) is satisfied. In this context we will often refer to the choice of a chamber as a \emph{closed BPS chamber}. The (instanton part of the) closed topological string partition function entering the above expression is given by \cite{GV-I,GV-II}
$$
\cZ_{top}(Q) = M(q_s)^{\chi/2} \prod_{l=1}^{\infty} \prod_{\beta>0, m} (1 - Q^{\beta} q_s^{m+l})^{l N^m_{\beta}},
$$
where $M(q_s) = \prod_l (1-q_s^l)^{-l}$ is the MacMahon function and $\chi$ is the Euler characteristic of the Calabi-Yau manifold. 

To be more precise, an identification as a topological string partition function or its square arises if $R>0$ in (\ref{Zcentral}). Because $R$ arises just as a multiplicative factor in (\ref{Zcentral}), degeneracies depend only on its sign. Therefore another extreme case corresponds to negative $R$ and $B$ sufficiently small, when only a single D6-brane contributes to the partition function
\be
\widetilde{\cZ}(R<0,0<B<<1) = 1.        \label{chamber-singleD6}
\ee
More generally, for $R<0$, BPS generating functions often (but not always) take finite form. 

In what follows we denote BPS generating functions in chambers with positive $R$ by $\cZ$, and in chambers with negative $R$ by $\widetilde{\cZ}$ (and often omit the subscript BPS). Topological string partition functions will be denoted by $\cZ_{top}$, while generating functions of melting crystals by ordinary $Z$.

\bigskip  

The above structure can be generalized by including in the initial D6-D2-D0 configuration additional D4-branes wrapping lagrangian cycles in the internal Calabi-Yau manifold and extending in two space-time dimensions \cite{AY-owc,DSV,openWalls}. For simplicity we consider a system with a single D4-brane wrapping a lagrangian cycle. There are now additional BPS states in two remaining spacetime dimensions arising from open D2-branes ending on these D4-branes. Their net degeneracies $N_{s,\beta,\gamma}$ are characterized, firstly, by the $SO(2)$ spin $s$ whose origin is most clearly seen from the M-theory perspective \cite{OV99,LMV2000}. Secondly, they depend on two-cycles $\beta$ wrapped by open M2-branes, as well as one-cycles $\gamma$ on which these M2-branes can end.\footnote{In case of $N$ D4-branes wrapping the same lagrangian cycle, these states would additionally arise in representations $R$ of $U(N)$ \cite{OV99}. In case of a single brane this reduces to $U(1)$, and such a dependence can be reabsorbed into a parameter specifying a choice of $\gamma$.}

Lifting this system to M-theory we obtain a background of the form $\textrm{Taub-NUT} \, \times \,  \textrm{Calabi-Yau} \, \times \, S^1$, with the additional D4-brane promoted to M5-brane. This M5-brane wraps the lagrangian submanifold $L$ inside Calabi-Yau, the time circle $S^1$, and $\mathbb{R}_+ \times S^1_{TN}$ inside the Taub-NUT space. 
A part of this lagrangian $L$ is a torus $T^2 = S^1_{TN} \times S^1$, which will lead to some modular properties of the BPS counting functions: this modularity will be manifest in one chamber, where the open topological string amplitude will be completed to the product of $\theta$ functions. This M5-brane also breaks the $SO(4)$ spatial symmetry down to $SO(2)\times SO(2)'$. We denote the spins associated to both $SO(2)$ factors respectively by $\sigma$ and $\sigma'$, and the degeneracies of particles with such spins by $N^{\sigma,\sigma'}_{\beta,\gamma}$. In addition to closed K{\"a}hler parameters $Q=e^{-T}$, let us also introduce open ones related to discs wrapped by M2-branes $z=e^{-d}$. The real and imaginary parts of $T$ encode respectively the sizes of two-cycles $\beta$ and the value of the $B$-field through them. The real and imaginary parts of $d$ encode respectively sizes of the discs and holonomies of the gauge fields around them. Similarly as in the closed string case, to get non-trivial ensemble of mutually supersymmetric states, we set the real parts of $T$ and $d$ to zero, and consider non-trivial imaginary parts.

From the M-theory perspective we are interested in counting the net degeneracies of M2-branes ending on this M5-brane
$$
N_{\sigma,\beta,\gamma} = \sum_{\sigma'} (-1)^{\sigma'} N^{\sigma,\sigma'}_{\beta,\gamma}.
$$ 
In the remaining three-dimensional space, in the $R\to\infty$ limit, the M2-branes ending on the M5-brane are represented by a gas of free particles. These particles have excitations in $\mathbb{R}^2$ which we identify with the $z_1$-plane. To each such BPS particle, similarly as in the closed string case discussed above and in \cite{WallM,DVV-Mduality}, we can associate a holomorphic field
\be
\Phi(z_1) = \sum_{l} \alpha_{l} z_1^{l}  \label{phi-z1}.
\ee 
The modes of this field create states with the intrinsic spin $s$ and the orbital momentum $l$ in the $\mathbb{R}^2$ plane. The derivation of the BPS degeneracies relies on the identification of this total momentum $\sigma +l$ in the $R\to \infty$ limit, with the Kaluza-Klein modes associated to the rotations along $S^1_{TN}$ for the finite $R$, following the five-dimensional discussion in \cite{DVV-Mduality,connect4d5d}.  

The BPS generating functions we are after are given by a trace over the Fock space built by the oscillators of the second quantized field $\Phi(z_1)$, and restricted to the states which are mutually supersymmetric. In such a trace each oscillator from (\ref{phi-z1}) gives rise to one factor of the form $(1 - q_s^{\sigma+l-1/2} Q^{\beta} z^{\gamma} )^{\pm 1}$, where the exponent $\pm 1$ corresponds to the bosonic or fermionic character of the top component of the BPS state, 
\be
\cZ^{open}_{BPS} =  
 \prod_{\sigma,\beta,\gamma} \prod_{l=1}^{\infty} (1 - q_s^{\sigma+l-1/2} Q^{\beta} z^{\gamma} )^{N_{\sigma,\beta,\gamma}}\, |_{chamber}, \label{Zbps-open}
\ee
where the product is over either both positive or both negative $(\beta,\gamma)$. The parameters $q,Q$ and $z$ specify the chamber structure: the restriction to a given $chamber$ is implemented by imposing the condition on a central charge, analogous to (\ref{Zpositive}), 
\be
q_s^{\sigma+l-1/2} Q^{\beta} z^{\gamma}  <  1.         \label{Zopen}
\ee
This condition in fact specifies a choice of both \emph{closed} and \emph{open} chambers. The walls of marginal stability between chambers correspond to subspaces where, for some oscillator, the above product becomes 1, and then the contribution from such an oscillator drops out from the BPS generating function.

Similarly as in the closed string case, the above degeneracies can be related to open topological string amplitudes, rewritten 
in \cite{OV99} in the form
$$
\cZ_{top}^{open} = \exp\Big( \sum_{n=1}^{\infty} \sum_{\sigma} \sum_{\beta,\gamma>0}  N_{\sigma,\beta,\gamma} \frac{q_s^{n \sigma} Q^{n \beta} z^{n\gamma} }{n(q_s^{n/2}-q_s^{-n/2})}   \Big)   ,  
$$
with integer Ooguri-Vafa invariants $N_{\sigma,\beta,\gamma}$.\footnote{In case of $N$ D4-branes wrapping a lagrangian cycle this structure is again more complicated, because the states in $\mathbb{R}^3$ arise in representations of $U(N)$ \cite{OV99}. This requires replacing the factor $z^{n\gamma}$ by the sum $\sum_R \Tr_R V^n$ of traces in all possible representations $R$ of this $U(N)$ of the matrix $V$ encoding holonomies of the gauge fields. For simplicity we restrict here to the simplest case.}
 This formula 
represents in fact a series of quantum dilogarithms
\be
L(z,q_s) = \exp\Big(\sum_{n>0} \frac{z^n}{n(q_s^{n/2} - q_s^{-n/2})} \Big) = \prod_{n=1}^{\infty} (1 - z q_s^{n-1/2}),    \label{qdilog}
\ee
and can be written in the product form 
\be
\cZ_{top}^{open}(Q,z) = \prod_{\sigma} \prod_{\beta,\gamma>0} \prod_{n=1}^{\infty} \Big(1 - Q^{\beta}z^{\gamma} q_s^{\sigma+n-1/2}  \Big)^{N_{\sigma,\beta,\gamma}}.   \label{Ztop-open}
\ee
Comparing with (\ref{Zbps-open}) we conclude that the BPS counting functions take form of the modulus square of the open topological string amplitude
\be
\cZ^{open}_{BPS} = \cZ_{top}^{open}(Q,z) \cZ_{top}^{open}(Q^{-1},z^{-1}) \, |_{chamber}.           \label{Zbps-Ztop2}
\ee

Similarly as in the closed string case, there are also a few particularly interesting chambers to consider. For example, in the extreme chamber corresponding to Im$\,T$, Im$\, d\to 0$, the trace is performed over the full Fock space and yields the modulus square of the open topological string partition function. In this case the quantum dilogarithms arise in pairs, which (using the Jacobi triple product identity) combine to the modular function $\theta_3/\eta$; in consequence the total BPS generating function is modular and expressed as a product of such functions.


\subsection{Crystal interpretation}      \label{ssec-crystals}

Closed BPS generating functions (\ref{cZ-chamber}) turn out to be generating functions of statistical models of crystals, when parameters relevant for both interpretations are appropriately matched. Physical reasons for such relations have been given in \cite{foam,ChuangJaff,OY1}, and mathematical interpretation arose from works \cite{MNOP-I,Szendroi,YoungBryan}. Such crystal interpretation arises also from the fermionic formulation \cite{ps-pyramid,NagaoVO}, as we will review below. These crystals, in a more intricate way \cite{openWalls}, encode also open BPS generating functions (\ref{Zbps-open}). However, before discussing details of all these constructions, in this introductory section we present crystal models for two simplest toric Calabi-Yau manifolds, i.e. $\C^3$ and conifold.

$\C^3$ is the simplest Calabi-Yau manifold. It has no compact two-cycles, so relevant BPS states are bound states of arbitrary number of D0-branes with a single D6-brane wrapping entire $\C^3$. Their generating function is therefore expressed in terms of a single parameter $q_s=e^{-g_s}$. There is just a single non-zero Gopakumar-Vafa invariant $N^0_{\beta=0}=-1$, and as follows from (\ref{cZ-chamber}) this generating function coincides with the so called MacMahon function
\be
\cZ_{BPS} = \prod_{l=1}^{\infty}  \frac{1}{(1-q_s^l)^l} = M(q_s).     \label{macmahon}
\ee

\begin{figure}[htb]
\begin{center}
\includegraphics[width=0.4\textwidth]{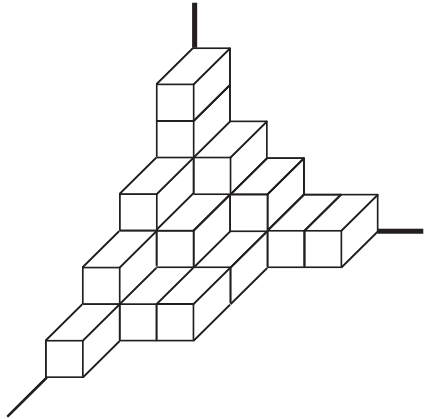} 
\begin{quote}
\caption{\emph{Plane partitions represent melting crystal configurations of $\C^3$.}} \label{fig-C3plane}
\end{quote}
\end{center}
\end{figure}

On the other hand, the MacMahon function is a generating function of plane partitions, i.e. three-dimensional generalization of Young diagrams. These plane partitions represent the simplest three-dimensional crystal model, namely they can be identified with stacks of unit cubes filling the positive octant of $\mathbb{R}^3$ space, as shown in fig. \ref{fig-C3plane}. A unit cube located in position $(I,J,K)$ can evaporate from this crystal only if all other cubes with coordinates $(i\leq I,j\leq J,k\leq K)$ are already missing. A plane partition $\pi$ is weighted by the number of boxes it consists of $|\pi|$, with a weight $q$ associated to a single box, so indeed
$$
Z = \sum_{\pi} q^{|\pi|} = \sum_{l=0}^{\infty} p(l) q^l = 1 + q + 3q^2 + 6q^3 + 13q^4 + \ldots = M(q),
$$
where $p(l)$ is the number of plane partitions which consist of $l$ cubes. Therefore plane partition generating function coincides with the BPS counting function $Z = \cZ_{BPS}$ when a simple identification
\be
q_s = q   \label{qs-q}
\ee
is made. From (\ref{ZBPS-Ztop}) it follows that the topological string partition function for $\C^3$ is given by the square root of the MacMahon function
\be
\cZ_{top} = M(q_s)^{1/2},   \label{MacMahon-C3}
\ee
which is indeed true. The relevance of the MacMahon function for $\C^3$ geometry was noticed for the first time in \cite{GV-I}, and a statistical model interpretation of this result was proposed in \cite{ok-re-va}. We also discuss its appearance from the matrix model viewpoint at the end of section \ref{ssec-matrix-conifold}.

\bigskip



\begin{figure}[htb]
\begin{center}
\includegraphics[width=0.9\textwidth]{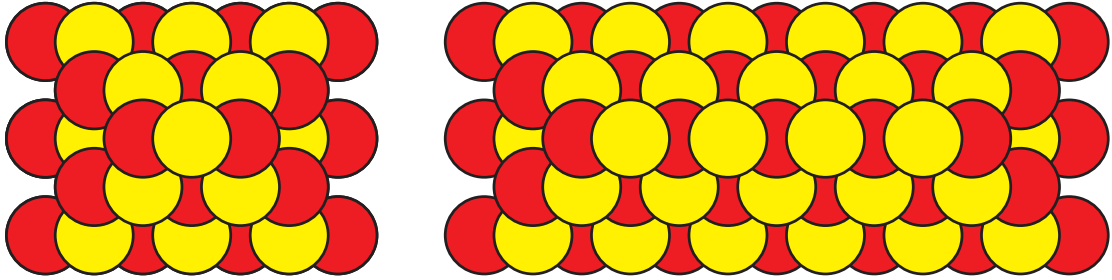} 
\begin{quote}
\caption{\emph{Infinite pyramids with one and four balls in the top row, with generating functions given respectively by $Z_0^{pyramid}$ and $Z_3^{pyramid}$.}} \label{fig-pyramids-coni-inf}
\end{quote}
\end{center}
\end{figure}

\begin{figure}[htb]
\begin{center}
\includegraphics[width=0.6\textwidth]{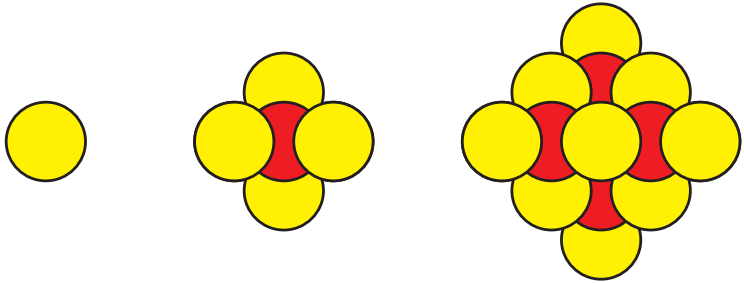} 
\begin{quote}
\caption{\emph{Finite pyramids with $m=1,2,3$ stones in the top row (respectively left, middle and right), whose generating functions are given by $\widetilde{Z}_{m+1}^{pyramid}$ (note that $\widetilde{Z}_1^{pyramid}=1$ corresponds to an empty pyramid corresponding to the pure D6-brane).}} \label{fig-pyramids-coni-fin}
\end{quote}
\end{center}
\end{figure}

The conifold provides another simple, yet non-trivial example of toric Calabi-Yau manifold. It consists of two $\C^3$ patches glued into $\mathcal{O}(-1)\oplus\mathcal{O}(-1)\to\mathbb{P}^1$, and it has one K{\"a}hler class representing $\mathbb{P}^1$, parametrized by $Q=e^{-T}$. This class can be wrapped by D2-branes, which bind with D0-branes to an underlying D6-brane and give rise to BPS states in low energy theory. In this case there is already a non-trivial structure of chambers and walls, which was analyzed in \cite{JaffMoore,ChuangJaff,Szendroi,NagaoNakajima}. This structure is consistent with M-theory derivation discussed in section \ref{ssec-M}. The generating functions of D6-D2-D0 bound states are parametrized by $Q$ and $q_s$, and therefore corresponding crystal models consist of two-colored three-dimensional partitions. The K{\"a}hler moduli space consists of several infinite countable sets of chambers, and in each chamber relevant crystal configurations take form of so called pyramid partitions. These partitions are infinite or finite (respectively for positive and negative $R$ in (\ref{Zcentral})) and their size depends on the value of the $B$-field. This size changes discretely and the pyramid is enlarged when the value of the $B$-field crosses integer numbers, which changes the chamber in the moduli space, as explained in previous subsection. Examples of such infinite pyramid partitions are given in fig. \ref{fig-pyramids-coni-inf}, and finite ones in fig. \ref{fig-pyramids-coni-fin}. 

To write down explicitly BPS generating functions for the conifold in various chambers we can take advantage of their relation to the topological string amplitude (\ref{ZBPS-Ztop}). The topological string partition function in this case reads
\be
\cZ^{conifold}_{top} (Q) = M(q_s)  \prod_{k\geq 1}(1 - Q q_s^k)^k,   \label{Ztop-conifold}
\ee
with the MacMahon function defined in (\ref{macmahon}).
From this topological string partition function we can read off Gopakumar-Vafa invariants \cite{GV-I,GV-II}
$$
N^0_{\beta=0} = -2, \qquad \qquad N^0_{\beta = \pm 1} = 1.
$$

Using the relation (\ref{ZBPS-Ztop}) we can now present conifold closed BPS generating functions in several sets of chambers. In the first set of chambers we consider $R>0$ and positive $B\in ]n,n+1 [$ (for $n\geq 0$). Firstly, for small $B$, there is so-called non-commutative chamber discussed first by Szendroi \cite{Szendroi}, which corresponds to $n=0$. In this case the pyramid crystal has just a single ball in the top row, as in the left panel in fig. \ref{fig-pyramids-coni-inf}, and the BPS generating function is given by the square of the topological amplitude. On the other hand, for large $B$, i.e. $n\to\infty$, we reach commutative chamber in which the length of the top row extends to infinity. In this case the BPS generating function agrees, up to a single factor of MacMahon function, with the topological string amplitude. In between there are chambers with $n+1$ balls in the top row, for which 
\be
\cZ_{n}^{conifold} = M(q_s)^2 \prod_{k\geq 1}(1 - Q q_s^k)^k  \prod_{k\geq n+1}(1 - Q^{-1} q_s^k)^{k} .   \label{ZnDT}
\ee
These BPS generating functions are related to pyramid generating functions with two colors $q_0$ and $q_1$ upon the identification (which generalizes (\ref{qs-q}) in $\mathbb{C}^3$ case)
$$
\cZ_n^{conifold} \ \textrm{chambers}:  \quad q_s = q_0 q_1, \quad  Q = -q_s^n q_1.
$$
Indeed, with this identification the above counting functions agree with those of two-colored pyramid crystals with 
$n+1$ yellow balls in its top row 
\be
Z_n^{pyramid}(q_0,q_1) = M(q_0 q_1)^2 \prod_{k\geq n+1}(1 + q_0^k q_1^{k+1})^{k-n} \prod_{k\geq 1}(1 + q_0^k q_1^{k-1})^{k+n}.   \label{Zn}
\ee

In the second set of chambers we have $R<0$ and positive $B\in ]n-1,n [$ (for $n\geq 1$). It extends between the core region with a single D6-brane (\ref{chamber-singleD6}) and the chamber characterized by so-called Pandharipande-Thomas invariants (for the flopped geometry, or equivalently for anti-M2-branes). The BPS generating functions read
\be
\widetilde{\cZ}_{n}^{conifold}  = \prod_{j=1}^{n-1} \big(1 - \frac{q_s^j}{Q} \big)^j.  \label{ZZnDT}
\ee 
The corresponding statistical models were shown in \cite{ps-pyramid,ChuangJaff,NagaoNakajima} to correspond to finite pyramids with $n-1$ stones in the top row, as shown in figure \ref{fig-pyramids-coni-fin}. In this case the generating functions of such partitions are equal to
\be
\widetilde{Z}_{n}^{pyramid}(q_0,q_1) = \prod_{j=1}^{n-1} (1 + q_0^{n-j} q_1^{n-j-1} )^j.  \label{ZZn}
\ee
The equality $\widetilde{\cZ}_{n}^{conifold}  \equiv \widetilde{Z}_n^{pyramid} $ arises upon an identification
$$
\widetilde{\cZ}_n^{conifold}  \ \textrm{chambers}: \quad q_s^{-1} = q_0 q_1, \quad Q = -q_s^{n} q_1.
$$
There are two other sets of chambers characterized by the negative value of the $B$-field, for which BPS generating functions are completely analogous to those given above. 

\bigskip

Above we presented just the simplest examples of crystal models. Using fermionic formulation presented below one can find other crystal models for arbitrary toric geometry without compact four-cycles. Let us also mention that those models can be equivalently expressed in terms dimers. In particular the operation of enlarging the crystal, as in the conifold pyramids, corresponds to so called \emph{dimer shuffling} \cite{YoungBryan}. Dimers are also closely related to a formulation using quivers and associated potentials, which underlies physical derivations in \cite{ChuangJaff,OY1}. 


\section{A little background -- free fermions and matrix models}    \label{sec-background}

In this section we introduce some mathematical background on which the main results presented in this review rely. In section \ref{ssec-toric} we start with a brief presentation of toric Calabi-Yau manifolds and introduce the notation which we use in what follows. In section \ref{ssec-fermion} we introduce free fermion formalism. In section \ref{ssec-matrix} we introduce basics of matrix model formalism. Our presentation is necessarily brief and for more detailed introduction we recommend many excellent reviews on each of those topics. 


\subsection{Toric Calabi-Yau three-folds}   \label{ssec-toric}

\begin{figure}[htb]
\begin{center}
\includegraphics[width=0.9\textwidth]{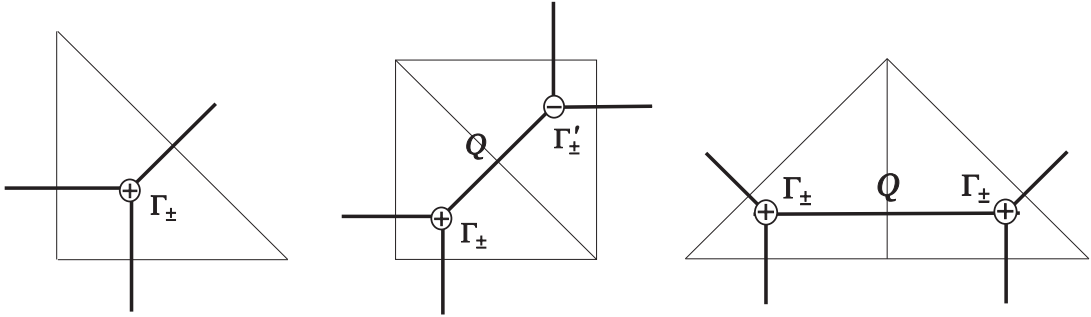} 
\begin{quote}
\caption{\emph{Toric graphs for $\C^3$, conifold and resolution of $\C^3/\Z_2$.}} \label{fig-strips}
\end{quote}
\end{center}
\end{figure}

Some introductory material on toric Calabi-Yau manifolds, from the perspective relevant for mirror symmetry and topological string thoery, can be found e.g. in \cite{MirrorSymmetry}. In this section our presentation is brief and mainly sets up the notation. Toric Calabi-Yau three-folds arise as the quotient of $\C^{\kappa+3}$, possibly with a discrete set of points deleted, by the action of $(\C^*)^{\kappa}$ with certain weights. The simplest toric three-fold is $\C^3$, which corresponds to the trivial choice $\kappa=0$. The resolved conifold, which we already discussed in section \ref{ssec-crystals}, corresponds to $\kappa=1$ and a choice of weights $(1,1,-1,-1)$, which represent a local bundle $\mathcal{O}(-1)\oplus\mathcal{O}(-1)\to\mathbb{P}^1$. The structure of each toric three-fold can be encoded in a two-dimensional diagram built from trivalent vertices. Finite intervals joining two adjacent vertices represent local $\mathbb{P}^1$ neighborhood inside the manifold. Equivalently one can consider dual graphs. Examples of toric diagrams and their duals for $\C^3$, conifold and resolution of $\C^3/\Z_2$ singularity are given in fig. \ref{fig-strips} (the notation $\Gamma_{\pm}$ at each vertex will be explained in what follows).

A closed loop in a toric diagram represents a compact four-cycle in the geometry. As follows from the reasoning in section \ref{ssec-M}, in the context of BPS counting we are forced to restrict considerations to manifolds which do not have such four-cycles. Apart from a few special cases, there is an infinite class of such geometries whose dual diagrams arise from a triangulation, into triangles of area $1/2$, of a long rectangle or a strip of height 1. A toric diagram arises as a dual graph to such a triangulation. From each vertex in such a toric diagram one vertical line extends to infinity and crosses either the upper or the lower edge of the strip. Two such consecutive lines can emanate either in the same or in the opposite direction, respectively when they are the endpoints of an interval representing $\mathbb{P}^1$ with local $\mathcal{O}(-2)\oplus\mathcal{O}$ or $\mathcal{O}(-1)\oplus\mathcal{O}(-1)$ neighborhood. Example of a generic diagram of this kind is shown in fig. \ref{fig-strip} below. 

Let us denote independent $\mathbb{P}^1$'s, starting from the left end of the strip, from 1 to $N$, and introduce corresponding K{\"a}hler parameters $Q_i=e^{-T_i}$, $i=1,\ldots,N$. Moreover to each toric vertex we associate a type $t_i=\pm 1$, so that $t_{i+1}=t_i$
if the local neighborhood of $\mathbb{P}^1$ (represented by an interval between vertices $i$ and $i+1$) is $\mathcal{O}(-2)\oplus\mathcal{O}$; if this neighborhood is of $\mathcal{O}(-1)\oplus\mathcal{O}(-1)$ type, then $t_{i+1}=-t_i$. The type of the first vertex we fix as $t_1=+1$. In figures \ref{fig-strips} and \ref{fig-strip} these types are denoted by $\oplus$ and $\ominus$. The types $t_i$ will be used much in the construction of fermionic states in section \ref{ssec-triangulate}.

As explained in section \ref{ssec-M}, the BPS generating functions can be expressed in terms of (the instanton part) of topological string amplitudes. For the above class of geometries, arising from a triangulation of a strip, these amplitudes read
\be
\cZ_{top}(Q_i) =  M(q_s)^{\frac{N+1}{2}} \prod_{l=1}^{\infty} \prod_{1\leq i < j \leq N+1} \Big(1- q_s^l\, (Q_i Q_{i+1}\cdots Q_{j-1}) \Big)^{-(t_i t_j) l}.   \label{Ztop-strip}
\ee


\subsection{Free fermion formalism}       \label{ssec-fermion}

Formalism of free fermions in two dimensions is well known \cite{jimbo-miwa,macdonald} and ubiquitous in literature on topological strings and crystal melting \cite{ok-re-va,YoungBryan,ps-phd,YoungBryan,adkmv}. The main purpose of this section is therefore to set up the notation which we will follow in the remaining parts of this paper. 

\begin{figure}[htb]
\begin{center}
\includegraphics[width=0.4\textwidth]{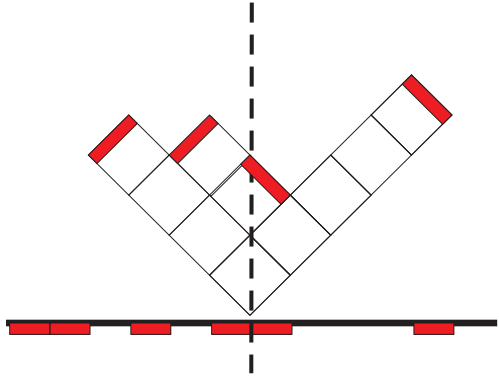} 
\begin{quote}
\caption{\emph{Relation between Young diagrams and states in the Fermi sea.}} \label{fig-young-holes}
\end{quote}
\end{center}
\end{figure}

The states in the free fermion Fock space are created by the (anti-commuting) modes of the fermion field 
$$
\psi(z) = \sum_{k\in\mathbb{Z}} \psi_{k+1/2} z^{-k-1}, \qquad \psi^*(z) = \sum_{k\in\mathbb{Z}} \psi^*_{k+1/2} z^{-k-1},  \qquad \{ \psi_{k+1/2},\psi^*_{-l-1/2} \} = \delta_{k,l}
$$ 
on the vacuum state $|0\rangle$. There is one-to-one map between such fermionic states 
$$
|\mu\rangle = \prod_{i=1}^d \psi^*_{-a_i-1/2} \psi_{-b_i-1/2} |0\rangle, \qquad \textrm{with} \quad  a_i = \mu_i-i,\ b_i =\mu^t_i-i,
$$
and two-dimensional partitions $\mu=(\mu_1,\mu_2,\ldots \mu_l)$, as shown in fig. \ref{fig-young-holes}. The modes $\alpha_m$ of the bosonized field $\partial \phi = :\psi(z)\psi^*(z):$ satisfy the Heisenberg algebra $[\alpha_m,\alpha_{-n}] = n \delta_{m,n}.$

We introduce  vertex operators
\be
\G_{\pm}(x) = e^{\sum_{n>0} \frac{x^n}{n}\alpha_{\pm n}}, \qquad \qquad \G'_{\pm}(x) = e^{\sum_{n>0} \frac{(-1)^{n-1}x^n}{n}\alpha_{\pm n}},     \label{vertexoperators}
\ee
which act on fermionic states $|\mu\rangle$ corresponding to partitions $\mu$ as \cite{jimbo-miwa,macdonald,YoungBryan}
\bea
\G_-(x) |\mu\rangle  =  \sum_{\lambda \succ \mu} x^{|\lambda|-|\mu|}|\lambda\rangle,  & &\qquad \qquad  
\G_+(x) |\mu\rangle  =  \sum_{\lambda \prec \mu} x^{|\mu|-|\lambda|}|\lambda\rangle,   \label{Gmu}   \\
\G'_-(x) |\mu\rangle =  \sum_{\lambda^t \succ \mu^t} x^{|\lambda|-|\mu|}|\lambda\rangle, & & \qquad \qquad
\G'_+(x) |\mu\rangle  =  \sum_{ \lambda^t \prec \mu^t} x^{|\mu|-|\lambda|}|\lambda\rangle.   \label{Gprimmu}
\eea
The interlacing relation $\prec$ between partitions is defined as
\be
\lambda \succ \mu  \qquad  \Leftrightarrow \qquad  \lambda_1 \geq \mu_1 \geq \lambda_2 \geq \mu_2 \geq \lambda_3 \geq \ldots.  \label{interlace}
\ee

The operator $\G'$ is the inverse of $\G$ with negative argument. These operators satisfy commutation relations
\bea
\G_+(x) \G_-(y) & = & \frac{1}{1-xy} \G_-(y) \G_+(x),   \label{GplusGminus}\\
\G'_+(x) \G'_-(y) & = & \frac{1}{1-xy} \G'_-(y) \G'_+(x),   \\
\G'_+(x) \G_-(y) & = & (1+xy) \G_-(y) \G'_+(x),   \\
\G_+(x) \G'_-(y) & = & (1+xy) \G'_-(y) \G_+(x).
\eea

We also introduce various colors $q_g$ and the corresponding operators $\widehat{Q}_g$ (a hat is to distinguish them from K{\"a}hler parameters $Q_i$)
\be
\widehat{Q}_g|\lambda\rangle = q_g^{|\lambda|}|\lambda\rangle.     \label{Qcolor}
\ee
These operators commute with vertex operators up to rescaling of their arguments
\bea
\G_+(x) \widehat{Q}_g = \widehat{Q}_g \G_+(x q_g), & & \qquad \qquad
\G'_+(x) \widehat{Q}_g = \widehat{Q}_g \G'_+(x q_g), \\
\widehat{Q}_g \G_-(x) = \G_-(x q_g) \widehat{Q}_g, & & \qquad \qquad 
\widehat{Q}_g \G'_-(x) = \G'_-(x q_g) \widehat{Q}_g.    \label{Qcommute}
\eea


\subsection{Matrix models}     \label{ssec-matrix}

In matrix model theory, or theory of random matrices, one is interested in properties of various ensembles of matrices. Excellent 
reviews of random matrix theory can be found e.g. in \cite{2dRandomMatrices} or, in particular in the context of topological string theory, in \cite{Marino}. In matrix model theory one typically considers partition functions of the form
\be
Z =  \int \mathcal{D}U \prod_{\alpha} e^{-\frac{1}{g_s} \textrm{Tr}\, V(U)},   \label{Zmatrix}
\ee
where $V=V(U)$ is a matrix potential, and $\mathcal{D}U$ is a measure over a set of matrices of interest $U$ of size $N$. Typically it is not possible to perform the above integral, however special techniques allow to determine its formal $1/N$ expansion. These techniques culminated with the formalism of the topological expansion of Eynard and Orantin \cite{EO} which, in principle, allows to determine entire $1/N$ expansion of the partition function recursively. This solution is determined by the behavior of matrix eigenvalues, whose distribution among the minima of the potential, in the continuum limit, determines one-dimensional complex curve, so-called spectral curve. The spectral curve is also encoded in the leading $1/N$ expansion of the so-called resolvent, which is defined as the expectation value $\omega(x) = \langle \textrm{Tr} \frac{1}{x-U} \rangle$ computed with respect to the measure (\ref{Zmatrix}). 

In the context of BPS counting and topological strings, unitary ensembles of matrices of infinite size arise. In this case the matrix model simplifies to the integral over eigenvalues $u_{\alpha}$, with a measure which takes form of the unitary Vandermonde determinant
$$
\mathcal{D}U=\prod_{\alpha} du_{\alpha} \,\prod_{\alpha<\beta}|z_{\alpha}-z_{\beta}|^2,\qquad \qquad z_{\alpha}=e^{iu_{\alpha}}.
$$
The issue of infinite matrices is a little subtle, however it can be taken care of by considering matrices of large but finite size $N$, and subsequently taking $N\to\infty$ limit. For finite $N$ one can find the resolvent, and in consequence the spectral curve, using a standard technique of so-called Migdal integral. This requires redefining $V$ to the standard Vandermonde form \cite{Marino,OSY}, as well as introducing 't Hooft coupling $T$  
\be
V\to V+T\log z,\qquad \qquad T=Ng_s.   \label{measure-T}   
\ee
The form of the Migdal integral depends on the number of cuts into which eigenvalues condense in large $N$ limit, and this number of cuts determines the genus of the spectral curve. In our context only single-cut situations will arise, for which the spectral curve has genus zero. In this case the Migdal integral determines the resolvent as
\be
\omega(p) = \frac{1}{2T} \oint \frac{dz}{2\pi i} \frac{\partial_z V(z)}{p-z}\frac{\sqrt{(p-a)(p-b)}}{\sqrt{(z-a)(z-b)}},  \label{Migdal}
\ee
so that the integration contour encircles counter-clockwise the endpoints of the cut $a$ and $b$. A proper asymptotic behavior of the resolvent is imposed by the condition
\be
\lim_{p\to\infty} \omega(p) = \frac{1}{p}.    \label{Migdal-norm}
\ee
Then the spectral curve is determined as a surface on which the resolvent is unambiguously defined, i.e. it is given by an (exponential) rational equation automatically satisfied by $p$ and $\omega(p)$. There is also an important consistency condition for the resolvent: when computed on the opposite sides of the cut $\omega(p)_{\pm}$, it is related to the potential as
\be
\omega_+(p) + \omega_-(p) =\frac{\partial_pV(p)}{T}.       \label{omegaPM}
\ee
On the other hand, a difference of these values of the resolvent on both sides of the cut provides eigenvalue density
\be
\rho(p) = \omega_+(p) - \omega_-(p) .       \label{rho-omega}
\ee

\begin{figure}[htb]
\begin{center}
\includegraphics[width=0.8\textwidth]{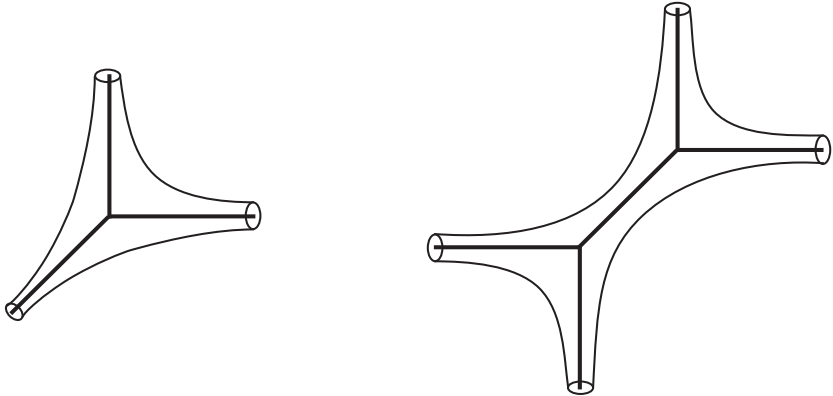} 
\begin{quote}
\caption{\emph{Toric diagrams for $\C^3$ and conifold and corresponding mirror curves.}} \label{fig-mirrorcurve}
\end{quote}
\end{center}
\end{figure}

It has been observed in several contexts that topological strings on toric manifolds can be related to matrix models, whose spectral curves take form of the so-called mirror curves. Mirror curves arise for manifolds which are mirror to toric Calabi-Yau manifolds \cite{MirrorSymmetry,adkmv}. For toric manifolds, their mirror manifolds are determined by the following equation embedded in four-dimensional complex space
$$
z_1 z_2 = H(x,y).
$$
The mirror curve is the zero locus of $H(x,y)$, i.e. it is given as $H(x,y)=0$. More precisely, $x,y$ are $\C^*$ variables, and it is often convenient to represent them in the exponential form $x=u^u$, $y=e^v$, with $u,v\in\C$. For example, for $\C^3$ and the conifold they take the following form
\be
H_{\C^3}(x,y) = x + y + xy = 0,   \qquad \quad  H_{conifold}(x,y) = x + y + x y + Qx^2= 0,    \label{mirrorcurves}
\ee
where $Q$ encodes the K{\"a}hler parameter of the conifold. Schematically mirror curves arise from thickening edges of the toric graphs, as shown in fig. \ref{fig-mirrorcurve}. 

One of the first relations between topological strings for toric manifolds and matrix models were encountered in \cite{mm-lens,lens-matrix}, where it was shown that the spectral curve of a unitary matrix model with a gaussian (i.e. quadratic) potential agrees with the above mirror curve $H_{conifold}(x,y)=0$ in (\ref{mirrorcurves}), with 't Hooft coupling $T=g_s N$ encoded in $Q=e^{-T}$. At the same time it was shown that the matrix model partition function reproduces the topological string partition function. More recently these ideas became important in view of the \emph{remodeling} conjecture \cite{BKMP,vb-ps}, which states that the solution to loop equations in the form found by Eynard and Orantin \cite{EO}, applied to the mirror curve, reproduces topological string partition functions. The method of \cite{EO} works for arbitrary curves, not necessarily originating from matrix models. Nonetheless, it is indeed possible to construct matrix models whose partition functions do reproduce topological string amplitudes, and whose spectral curves coincide with appropriate mirror curves    \cite{OSY,SzaboTierz-DT,eynard-planch,SW-matrix,matrix2star,2009betaMatrix,EynardTopological}. 

One of our aims is to provide matrix model interpretation of BPS counting. It is natural to expect such an interpretation in view of 
an intimate relation between BPS counting and topological string theory discussed in section \ref{ssec-M}, and 
the above mentioned relations between topological strings and matrix models. As we will see in what follows, there are indeed unitary matrix models which naturally arise in the context of BPS counting and its fermionic formulation. Among the others, our task will be to analyze them using the above mentioned Migdal method.

\section{Fermionic formulation of BPS counting functions}    \label{sec-BPSfermions}

Having introduced all the ingredients above, we are now ready to present fermionic formulation of BPS counting. To start with, in section \ref{ssec-C3example} we present the idea of such a formulation in the simplest example of $\C^3$. In section \ref{ssec-triangulate} we introduce a general fermionic formalism, and in section \ref{ssec-CrystalMelting} we provide its crystal interpretation. We illustrate the use of our formalism in section \ref{ssec-examplesFermions} revisiting $\C^3$ example, as well as in explicit case of $\C^3/\Z_N$, and conifold geometry.

\subsection{The idea and $\C^3$ example}    \label{ssec-C3example}

\begin{figure}[htb]
\begin{center}
\includegraphics[width=0.8\textwidth]{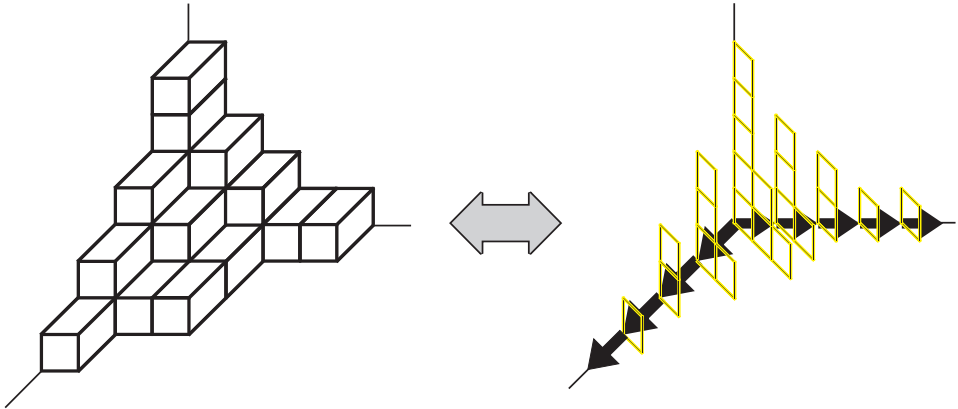} 
\begin{quote}
\caption{\emph{Slicing of a plane partition (left) into a sequence of interlacing two-dimensional partitions (right). A sequence of $\G_{\pm}$ operators in (\ref{C3-crystal}) which create two-dimensional partitions is represented by arrows inserted along two axes. Directions of arrows $\rightarrow$  represent interlacing condition $\succ$ on partitions. We reconsider this example from a new viewpoint in figure \ref{fig-C3new}.}} \label{fig-C3}
\end{quote}
\end{center}
\end{figure}

As explained in section \ref{ssec-crystals}, the generating function of bound states of D0-branes to a single D6-brane is given by the MacMahon function, and the corresponding crystal model takes form of the counting of plane partitions \cite{ok-re-va}. Let us slice each such plane partition by a set of parallel planes, as shown in fig. \ref{fig-C3}. In this way on each slice we obtain a two-dimensional partition $\mu$, and it is not hard to see that each two neighboring partitions satisfy the interlacing condition (\ref{interlace}). Recalling that such a condition arises if we apply $\G_{\pm}(1)$ operators (\ref{Gmu}) to partition states, we conclude that a set of all plane partitions can be built, slice by slice, by acting with infinite sequence of $\G_{\pm}(1)$ on the vacuum. To count each slice $\mu$ with appropriate weight $q^{|\mu|}$ we also need to apply weight operator $\widehat{Q}$ defined in (\ref{Qcolor}). Therefore the generating function of plane partitions can be represented as follows
\bea 
Z & = & \langle \Omega_+ | \Omega_- \rangle 
\equiv \langle 0 | \ldots \widehat{Q} \G_+(1) \widehat{Q} \G_+(1) \widehat{Q} \G_+(1) \ | \ \widehat{Q} \G_-(1) \widehat{Q} \G_-(1) \widehat{Q} \G_-(1) \widehat{Q} \ldots | 0 \rangle =  \nonumber \\
& = & \langle 0 | \ldots \G_+(q^2) \G_+(q) \G_+(1)  \G_-(q)  \G_-(q^2) \G_-(q^3) \ldots | 0 \rangle = \label{C3-crystal}  \\
& = & \prod_{l_1,l_2=1}^{\infty} \frac{1}{1-q^{l_1+l_2-1}} = M(q).   \nonumber
\eea
In the first line we implicitly introduced two states $\langle \Omega_+ |$ and $| \Omega_- \rangle$, defined by an infinite sequence of $\Gamma_{+}$ (respectively $\Gamma_-$) operators, interlaced with weight operators $\widehat{Q}$ and acting on the vacuum. To confirm that this correlator indeed reproduces the MacMahon function, the second line can be reduced to the final infinite product using commutation relations (\ref{GplusGminus}) and (\ref{Qcommute}). We can also represent insertions of $\G_{\pm}(1)$ operators graphically by arrows, so that the above computation can represented as in fig. \ref{fig-C3} (right). 

In what follows we present a formalism which allows to generalize this computation to a large class of chambers, for arbitrary toric geometry without compact four-cycles.


\subsection{Toric geometry and quantization}         \label{ssec-triangulate}

We wish to reformulate BPS counting in the fermionic language in a way in which we associate to each toric manifold a fermionic state, such that the BPS generating function can be expressed as an overlap of two such states, generalizing $\C^3$ case (\ref{C3-crystal}). At the same time the construction of such a fermionic state is supposed to encode the structure of the underlying crystal model (generalizing plane partitions in fig. \ref{fig-C3}). An important difference between $\C^3$ and other geometries is the existence of many K{\"a}hler moduli and correspondingly many chambers, for which BPS generating functions change according to wall-crossing formulas. To take care of these changes in the fermionic formalism we need to introduce special \emph{wall-crossing} operators.

\subsubsection{Toric geometry and fermionic operators}    \label{sssec-toric}

\begin{figure}[htb]
\begin{center}
\includegraphics[width=0.9\textwidth]{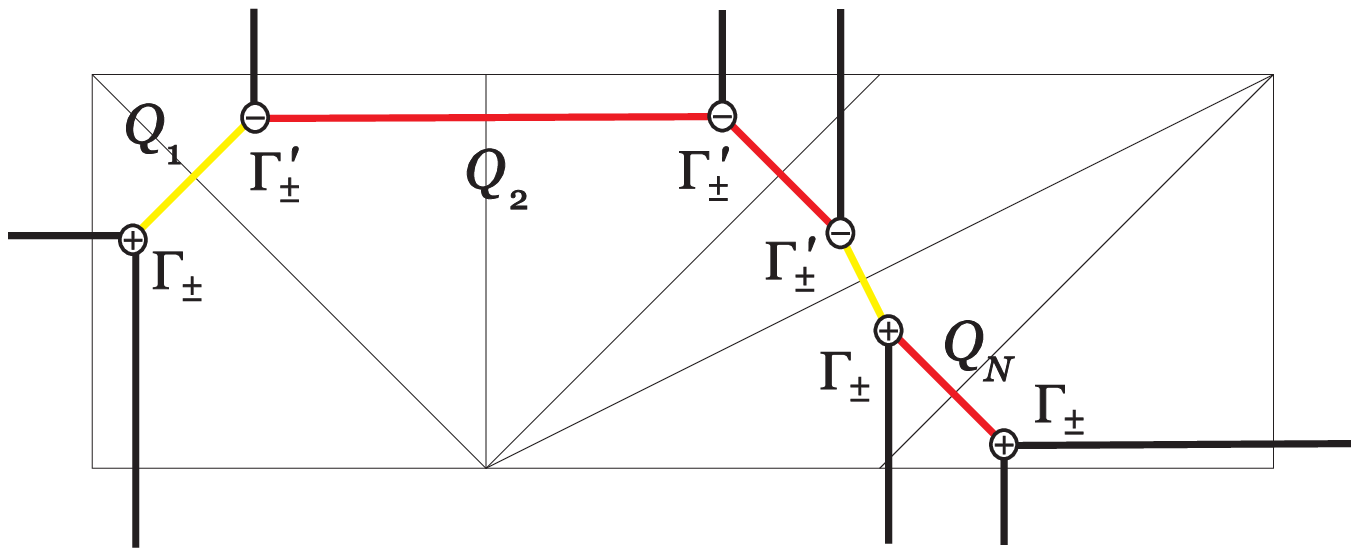} 
\begin{quote}
\caption{\emph{Toric Calabi-Yau manifolds represented by a triangulation of a strip. There are $N$ independent $\mathbb{P}^1$'s with K{\"a}hler parameters $Q_i=e^{-T_i}$, and $N+1$ vertices to which we associate $\G$ and $\G'$ operators represented respectively by $\oplus$ and $\ominus$ signs. Yellow intervals, which connect vertices with opposite signs, represent $\mathcal{O}(-1)\oplus\mathcal{O}(-1)\to \mathbb{P}^1$ local neighborhoods. Red intervals, which connect vertices with the same signs, represent $\mathcal{O}(-2)\oplus\mathcal{O}\to \mathbb{P}^1$ local neighborhoods. The first vertex on the left is chosen to be $\oplus$.}} \label{fig-strip}
\end{quote}
\end{center}
\end{figure}      

In what follows we use the notation introduced in section \ref{ssec-toric}, in particular to each vertex of the toric diagram we associate its type $t_i=\pm 1$, see also fig. \ref{fig-strip}. We start with a construction of fermionic states associated to a given toric Calabi-Yau manifold (without compact four-cycles). First we need to introduce several operators which are building blocks of such states. The structure of these operators is encoded in the toric diagram of a given manifold. Namely, these operators are given by a string of $N+1$ vertex operators $\G_{\pm}^{t_i}(x)$ (defined in (\ref{vertexoperators})) which are associated to the vertices of the toric diagram; the type $t_i$ determines the type of a vertex operator as
\be
\G_{\pm}^{t_i=+1}(x)=\G_{\pm}(x),\qquad \qquad \G_{\pm}^{t_i=-1}(x)=\G_{\pm}'(x).   \label{ti-GammaPM}
\ee
In addition the string of operators $\G_{\pm}^{t_i}(x)$ is interlaced with $N+1$ operators $\widehat{Q}_i$ representing colors $q_i$, for $i=0,1,\ldots,N$. Operators $\widehat{Q}_1,\ldots,\widehat{Q}_N$ are associated to $\mathbb{P}^1$ in the toric diagram, and there is an additional $\widehat{Q}_0$. We also define 
\be
\widehat{Q} = \widehat{Q}_0 \widehat{Q}_1\cdots \widehat{Q}_N,\qquad \qquad q = q_0 q_1 \cdots q_N.
\ee
Therefore the upper indices of $\G_{\pm}^{t_i}(x)$ and a choice of colors of the operators which we introduce below are specified by the data of a given toric manifold. As we will see, a sequence of lower indices $\pm$ is determined by the chamber we are going to consider.

Now we can associate several operators to a given toric manifold. Firstly we define
\be
\overline{A}_{\pm}(x) = \G_{\pm}^{t_1} (x) \widehat{Q}_1 \G_{\pm}^{t_2} (x) \widehat{Q}_2 \cdots \G_{\pm}^{t_N} (x) \widehat{Q}_N \G_{\pm}^{t_{N+1}} (x) \widehat{Q}_0.                  \label{Apm}
\ee
Commuting all $\widehat{Q}_i$'s  using (\ref{Qcommute}) we also define the following operators
\bea
A_+(x) & = & \widehat{Q}^{-1} \, \overline{A}_{+}(x) = \G_{+}^{t_1} \big(xq\big)  \G_{+}^{t_2} \big(\frac{xq}{q_1}\big) \G_{+}^{t_3} \big(\frac{xq}{q_1 q_2}\big) \cdots \G_{+}^{t_{N+1}} \big(\frac{xq}{q_1q_2\cdots q_N}\big), \label{Aplus} \\
A_-(x) & = & \overline{A}_{-}(x) \, \widehat{Q}^{-1} = \G_{-}^{t_1} (x)  \G_{-}^{t_2} (xq_1) \G_{-}^{t_3} (x q_1 q_2) \cdots \G_{-}^{t_{N+1}} (x q_1q_2 q_N). \label{Aminus}
\eea

In addition, we define the above mentioned \emph{wall-crossing operators}
\begin{align}
& \overline{W}_p(x) =  \Big(\G_{-}^{t_1} (x) \widehat{Q}_1 \G_{-}^{t_2} (x) \widehat{Q}_2 \cdots \G_{-}^{t_p} (x) \widehat{Q}_p\Big)\Big( \G_{+}^{t_{p+1}} (x) \widehat{Q}_{p+1} \cdots \G_{+}^{t_N} (x) \widehat{Q}_N \G_{+}^{t_{N+1}} (x) \widehat{Q}_0 \Big)  \label{Wp} \\
& \overline{W}_p'(x)  =  \Big(\G_{+}^{t_1} (x) \widehat{Q}_1 \G_{+}^{t_2} (x) \widehat{Q}_2 \cdots \G_{+}^{t_p} (x) \widehat{Q}_p\Big)\Big( \G_{-}^{t_{p+1}} (x) \widehat{Q}_{p+1} \cdots \G_{-}^{t_N} (x) \widehat{Q}_N \G_{-}^{t_{N+1}} (x) \widehat{Q}_0 \Big)   \label{Wpprim}  
\end{align}
Here the order of $\G$ and $\G'$ is the same as for $\overline{A}_{\pm}$ operators, and the difference is that now there are subscripts $\mp$ on first $p$ operators and $\pm$ on the remaining ones. 


We often use a simplified notation when the argument of the above operators is $x=1$
$$
\overline{A}_{\pm} \equiv \overline{A}_{\pm}(1), \qquad A_{\pm} \equiv A_{\pm}(1), \qquad \overline{W}_p \equiv \overline{W}_p(1), \qquad \overline{W}'_p \equiv \overline{W}'_p(1).
$$


\subsubsection{Fermionic formulation and quantization}         \label{ssec-FockState}

Above we associated operators $\overline{A}_{\pm}$ to each toric geometry with a strip-like toric diagram. From these operators we can build the following states in the Hilbert space of a free fermion $\cH$
\be
|\Omega_{\pm}\rangle  \in\cH,   \label{Omega-pm}
\ee
which we define as follows
\bea
\langle \Omega_+| & = & \langle 0 | \ldots \overline{A}_+(1) \overline{A}_+(1) \overline{A}_+(1) = \langle 0 | \ldots A_+(q^2) A_+(q) A_+(1),  \label{Omega-plus}  \\
| \Omega_- \rangle & = & \overline{A}_-(1) \overline{A}_-(1) \overline{A}_-(1) \ldots |0\rangle = A_-(1) A_-(q) A_-(q^2) \ldots |0\rangle .  \label{Omega-minus}
\eea
These states encode the full instanton part of the topological string amplitudes. Namely, as shown in \cite{ps-pyramid},
\be
Z = \langle \Omega_+ | \Omega_- \rangle    \label{Z-Omega}  
\ee
is equal to the BPS partition function $\cZ$ in the non-commutative chamber
\be
Z = \cZ \equiv |\cZ_{top}|^2 \equiv \cZ_{top}(Q_i) \cZ_{top}(Q_i^{-1}),    \label{Z-cZ}
\ee
where $\cZ_{top}(Q_i)$ is given in (\ref{Ztop-strip}). The above equality holds under the following identification between $q_i$ parameters (which enter the definition of $|\Omega_{\pm}\rangle$) and physical parameters $Q_i=e^{-T_i}$ and $q_s=e^{-g_s}$:
\be
q_i = (t_i t_{i+1}) Q_i,\qquad \qquad q_s = q \equiv q_0 q_1\cdots q_N.   \label{qQ}
\ee
We will provide a proof of (\ref{Z-Omega}) in section \ref{ssec-ncDT} in a more general setting of refined invariants. 


The states $|\Omega_{\pm}\rangle$ have non-trivial structure and encode the information about the non-commutative chamber. It turns out that the fermionic vacuum $|0\rangle$ itself also encodes some interesting information. We recall that there is another extreme chamber representing just a single BPS state represented by the D6-brane with no other branes bound to it. This multiplicity 1 can be understood as
\be
\widetilde{\cZ} = \widetilde{Z} = \langle 0 | 0 \rangle = 1,   \label{Ztilde-1}
\ee
and as we will see below, starting from this expression we can use wall-crossing operators to construct BPS generating functions in an infinite family of other chambers. 


\subsubsection{Other chambers and wall-crossing operators}   \label{ssec-WallCross}

In the previous section we associated to toric manifolds the states $|\Omega_{\pm}\rangle$, whose overlap reproduces the BPS generating function in the non-commutative chamber (\ref{Z-Omega}). Now we wish to extend this formalism to other chambers. As discussed in section \ref{ssec-M}, in a given chamber, the allowed bound states we wish to count must have positive central charge (\ref{Zcentral})
$$
Z(R,B) = \frac{1}{R}(n+\beta \cdot B) > 0.
$$
Firstly, the information about $R$ and $B$ must be encoded in the fermionic states which we wish to construct. It turns out that the choice of positive or negative $R$ is encoded in the choice of the ground state
\be
R>0 \quad \longrightarrow \quad |\Omega_{\pm} \rangle, \qquad \qquad R<0 \quad \longrightarrow \quad |0\rangle,   \label{DTPT-R}
\ee
which generalizes the extreme cases (\ref{Z-Omega}) and (\ref{Ztilde-1}).

On the other hand, the value of the field $B$ is encoded in the insertion of additional wall-crossing operators, such as those defined in (\ref{Wp}) and (\ref{Wpprim}). In particular these two types of operators are sufficient if we wish to consider only these chambers, which correspond to a flux of the $B$-field through only one, but arbitrary $\mathbb{P}^1$ in the manifold. For simplicity below we consider only this set of chambers. Denoting this $\mathbb{P}^1$ as $p$, it can be shown that insertion of $n$ copies of operators $\overline{W}_p$ or $\overline{W}'_p$ creates respectively $n$ positive or negative quanta of the flux through $p$'th $\mathbb{P}^1$. 

Therefore, schematically, the generating functions in chambers with $R>0$ read
\be
Z_n = \langle \Omega_+ | (\overline{W})^n  |\Omega_- \rangle,    \label{Z-Omega-Wn}  
\ee
and those with $R<0$ read
\be
\widetilde{Z}_n = \langle 0  | (\overline{W})^n  | 0 \rangle,    \label{Z-vacuum-Wn}  
\ee
with appropriate form of wall-crossing operators. More precisely, depending on the signs of $R$ and $B$, we need to consider four possible situations, which we present below. The proofs of all statements below, corresponding to these four situations, can be found in \cite{ps-pyramid}.


\begin{itemize}

\item

\textbf{Chambers with $R<0$, $B>0$}

Consider a chamber characterized by positive $R$ and positive $B$-field through $p$'th two-cycle
$$
R<0,\qquad \qquad  B\in ] n-1,n [ \quad \textrm{for}\quad 1\leq n \in \mathbb{Z}.
$$ 
The BPS partition function in this chamber contains only those factors which include $Q_p$ and it reads
$$
\widetilde{\cZ}_{n|p} = \prod_{i=1}^{n-1}  \prod_{s=1}^p \prod_{r=p+1}^{N+1} \Big(1 - \frac{q_s^{i}}{Q_s Q_{s+1}\cdots Q_{r-1}}  \Big) ^{-t_r t_s i}.
$$ 
This can be expressed as the expectation value of $n$ wall-crossing operators $\overline{W}_p$
\be
\widetilde{Z}_{n|p}  =  \langle 0| ( \overline{W}_p ) ^n |0 \rangle   = 
\widetilde{\cZ}_{n|p},
\label{ZRnegBpos-Results}  \\
\ee
under the following identification of variables
\be
Q_p = (t_p t_{p+1})q_p q_s^{n},\qquad \qquad Q_i = (t_i t_{i+1})q_i \quad \textrm{for}\quad i\neq p,\qquad \qquad q_s = \frac{1}{q}.  \label{ZRnegBpos-qQ-Results}
\ee
A special case of this result is the trivial generating function (\ref{Ztilde-1}) representing a single D6-brane.  



\item 
\textbf{Chambers with $R>0$, $B>0$}

In the second case we consider the positive value of $R$ and the positive flux through $p$'th $\mathbb{P}^1$
$$
R>0, \qquad \qquad B\in ]n,n+1 [ \quad \textrm{for} \quad   0\leq n \in \mathbb{Z}.
$$ 
Denote the BPS partition function in this chamber by $\cZ_{n|p}$. We find that the expectation value of $n$ wall-crossing operators $\overline{W}_p$ in the background of $|\Omega\rangle$ has the form 
\be
Z_{n|p} = \langle \Omega_+| ( \overline{W}_p ) ^n | \Omega_-\rangle  = M(1,q)^{N+1} \, Z^{(0)}_{n|p} \, Z^{(1)}_{n|p} \, Z^{(2)}_{n|p},  \label{ZRposBpos-Results}
\ee
where $Z^{(0)}_{n|p}$ does not contain any factors $(q_s\cdots q_{r-1})^{\pm 1}$ which would include $q_p$, while $Z^{(1)}_{n|p}$ contains all factors $q_s\cdots q_{r-1}$ which do include $q_p$, and $Z^{(2)}_{n|p}$ contains all factors $(q_s\cdots q_{r-1})^{-1}$ which also include  $q_p$:
\bea
Z^{(0)}_{n|p} & = & \prod_{l=1}^{\infty} \prod_{p \notin \overline{s,r+1} \subset \overline{1,N+1}} \Big(1 - (t_rt_s) \frac{q^{l}}{ q_s q_{s+1}\cdots q_{r-1}}  \Big) ^{-t_r t_s l}  \Big(1 - (t_rt_s) q^{l} q_s q_{s+1}\cdots q_{r-1} \Big) ^{-t_r t_s l} ,    \nonumber \\
Z^{(1)}_{n|p} & = & \prod_{l=1}^{\infty} \prod_{p \in \overline{s,r+1} \subset \overline{1,N+1}}   \Big(1 - (t_rt_s) q^{l+n} q_s q_{s+1}\cdots q_{r-1} \Big) ^{-t_r t_s l} ,    \nonumber \\
Z^{(2)}_{n|p} & = & \prod_{l=n+1}^{\infty} \prod_{p \in \overline{s,r+1} \subset \overline{1,N+1}} \Big(1 - (t_rt_s) \frac{q^{l-n}}{ q_s q_{s+1}\cdots q_{r-1}}  \Big) ^{-t_r t_s l} .     \nonumber 
\eea
We see that the identification of variables
\be
Q_p = (t_p t_{p+1})q_p q_s^{n},\qquad \qquad Q_i = (t_i t_{i+1})q_i \quad \textrm{for}\quad i\neq p,\qquad \qquad q_s = q  \label{ZRposBpos-qQ-Results}
\ee
reproduces the BPS partition function
\be
\cZ_{n|p} = Z_{n|p}.             \label{ZRposBpos-equal}
\ee
When no wall-crossing operator is inserted the change of variables reduces to (\ref{qQ}) and we get the non-commutative Donaldson-Thomas partition function (\ref{Z-cZ}), $Z_{0|p}=\cZ$.


\item

\textbf{Chambers with $R<0$, $B<0$}

Now we consider negative $R$ and negative $B$-field 
$$
R<0, \qquad \qquad   B \in ]-n-1,-n [ \quad \textrm{for} \quad 0\leq n \in \mathbb{Z}.
$$
For such a chamber the BPS partition function reads
$$
\widetilde{\cZ}'_{n|p} = \prod_{i=1}^{n}  \prod_{s=1}^p \prod_{r=p+1}^{N+1} \Big(1 - q_s^{i} Q_s Q_{s+1}\cdots Q_{r-1}  \Big) ^{-t_r t_s i}.
$$
Now we find the the expectation value of $n$ wall-crossing operators $\overline{W}'_p$ is equal to
\be
\widetilde{Z}'_{n|p} = \langle 0| ( \overline{W}'_p ) ^n | 0 \rangle  
= \widetilde{\cZ}'_{n|p},
\label{ZRnegBneg-Results}
\ee
under the change of variables
\be
Q_p = (t_p t_{p+1})q_p q_s^{-n},\qquad \qquad Q_i = (t_i t_{i+1})q_i \quad \textrm{for}\quad i\neq p,\qquad \qquad q_s = \frac{1}{q}.  \label{ZRnegBneg-qQ-Results}
\ee
Now an insertion of $\overline{W}_p$ has an interpretation of turning on a negative quantum of $B$-field, and the redefinition of $Q_p$ can be interpreted as effectively reducing $t_p$ by one unit of $g_s$. As already discussed
$$
\widetilde{Z}'_{0|p} = \langle 0 | 0 \rangle =  1
$$
represents a chamber with a single D6-brane and no other branes bound to it. 


\item 

\textbf{Chambers with $R>0$, $B<0$}

In the last case we consider positive $R$ and negative $B$
$$
R>0, \qquad \qquad  0>B\in ]-n,-n+1 [ \quad \textrm{for} \quad   1\leq n \in \mathbb{Z}.
$$
We denote the BPS partition function in this chamber by $\cZ'_{n|p}$. We find that the expectation value of $n$ operators $\overline{W}'_p$ in the background of $|\Omega_{\pm}\rangle$ has the form
\be
Z'_{n|p} = \langle \Omega_+| ( \overline{W}'_p ) ^n | \Omega_-\rangle  =   M(1,q)^{N+1} \, Z'^{(0)}_{n|p} \, Z'^{(1)}_{n|p} \, Z'^{(2)}_{n|p}  \label{ZRposBneg-Results}
\ee
where $Z'^{(0)}_{n|p}$ does not contain any factors $(q_s\cdots q_{r-1})^{\pm 1}$ which would include $q_p$,  $Z'^{(1)}_{n|p}$ contains all factors $q_s\cdots q_{r-1}$ which do include $q_p$, and $Z'^{(2)}_{n|p}$ contains all factors $(q_s\cdots q_{r-1})^{-1}$ which also include  $q_p$:
\bea
Z'^{(0)}_{n|p} & = & \prod_{l=1}^{\infty} \prod_{p \notin \overline{s,r+1} \subset \overline{1,N+1}} \Big(1 - (t_rt_s) \frac{q^{l}}{ q_s q_{s+1}\cdots q_{r-1}}  \Big) ^{-t_r t_s l}  \Big(1 - (t_rt_s) q^{l} q_s q_{s+1}\cdots q_{r-1} \Big) ^{-t_r t_s l} ,    \nonumber \\
Z'^{(1)}_{n|p} & = & \prod_{l=n}^{\infty} \prod_{p \in \overline{s,r+1} \subset \overline{1,N+1}}   \Big(1 - (t_rt_s) q^{l-n} q_s q_{s+1}\cdots q_{r-1} \Big) ^{-t_r t_s l} ,    \nonumber \\
Z'^{(2)}_{n|p} & = & \prod_{l=1}^{\infty} \prod_{p \in \overline{s,r+1} \subset \overline{1,N+1}} \Big(1 - (t_rt_s) \frac{q^{l+n}}{ q_s q_{s+1}\cdots q_{r-1}}  \Big) ^{-t_r t_s l} .     \nonumber 
\eea
Under the change of variables
\be
Q_p = (t_p t_{p+1})q_pq_s^{-n-1},\qquad \qquad Q_i = (t_i t_{i+1})q_i \quad \textrm{for}\quad i\neq p,\qquad \qquad q_s = q.  \label{ZRposBneg-qQ-Results}
\ee
this reproduces the BPS partition function
\be
\cZ'_{n|p} = Z'_{n|p}.    \label{ZRposBneg-equal}
\ee
We note that both $Z'_{1|p}$ with the above change of variables, as well as $Z_{0|p}$ given in (\ref{ZRposBpos-Results}) with a different change of variables in (\ref{ZRposBpos-qQ-Results}), lead to the same BPS generating function $\cZ$ which corresponds to the non-commutative Donaldson-Thomas invariants.

\end{itemize}


\subsection{Crystal melting interpretation}   \label{ssec-CrystalMelting}

In the previous section we found a free fermion representation of D6-D2-D0 generating functions.
The fermionic correlators which reproduce BPS generating functions automatically provide melting crystal interpretation of these functions \cite{ps-pyramid}, generalizing models of plane partitions (for $\mathbb{C}^3$) or pyramid partitions (for the conifold), presented in section  \ref{ssec-crystals}. These crystals are also equivalent to those found in \cite{NagaoVO,OY1}.


The crystal interpretation is a consequence of the fact that all operators used in the construction of states $|\Omega_{\pm}\rangle$, as well as the wall-crossing operators, are built just from vertex operators $\G_{\pm}$ and $\G_{\pm}$ with argument 1, and color operators $\widehat{Q}_i$. As follows from (\ref{Gmu}) and (\ref{Gprimmu}), insertion of these vertex operators is equivalent to the insertion of two-dimensional partitions satisfying interlacing, or transposed interlacing conditions. An (infinite) sequence of such interlacing partitions effectively builds up a three-dimensional crystal. A relative position of two adjacent slices is determined by a type of two corresponding vertex operators. On the other hand, insertions of color operators have an interpretation of coloring the crystal. The colors $\widehat{Q}_i$ appear in the same order in each composite operator, so these colors are always repeated periodically in the full correlators. Therefore, three-dimensional crystals are built of interlacing, periodically colored slices.

\begin{figure}[htb]
\begin{center}
\includegraphics[width=0.5\textwidth]{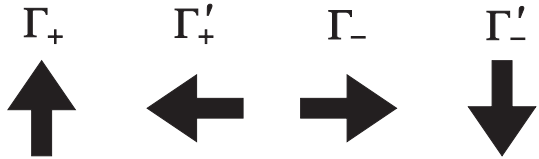} 
\begin{quote}
\caption{\emph{Assignment of arrows.}} \label{fig-arrows}
\end{quote}
\end{center}
\end{figure}

To get more insight about a geometric structure of a crystal it is convenient to introduce the following graphical representation. 
We associate various arrows to the vertex operators, as shown in fig. \ref{fig-arrows}. These arrows follow the order of the vertex operators in the fermionic correlators, and are drawn from left to right, or up to down (either of these directions is independent of the orientation of the arrow). Following the order of the vertex operators in a given correlator, and drawing a new arrow at the end of the previous one, produces a zig-zag path which represents a shape of the crystal. The coloring of the crystal is taken care of by keeping track of the order of $\widehat{Q}_i$ operators, and by drawing at the endpoint of each arrow a (dashed) line, rotated by 45$^\textrm{o}$, colored according to $\widehat{Q}_i$ which we come across. These lines represent two-dimensional slices in appropriate colors. In this way the corners of two-dimensional partitions arising from slicing of the crystal are located at the end-points of the arrows. The orientation of arrows represents the interlacing condition (i.e. arrows point from a larger to smaller partition). The interlacing pattern between two consecutive slices corresponds to the types of two consecutive arrows. Finally, the points from which two arrows point outwards represent those stones in the crystal, which can be removed from the initial, full crystal configuration. In fermionic correlators these points correspond to $\G^{t_i}_+$ followed by $\G^{t_j}_-$ operators. We illustrate this graphical construction in a few examples in the next section.

\subsection{Examples}   \label{ssec-examplesFermions}

\subsubsection{Revisiting $\mathbb{C}^3$}    \label{ssec-C3}

\begin{figure}[htb]
\begin{center}
\includegraphics[width=0.8\textwidth]{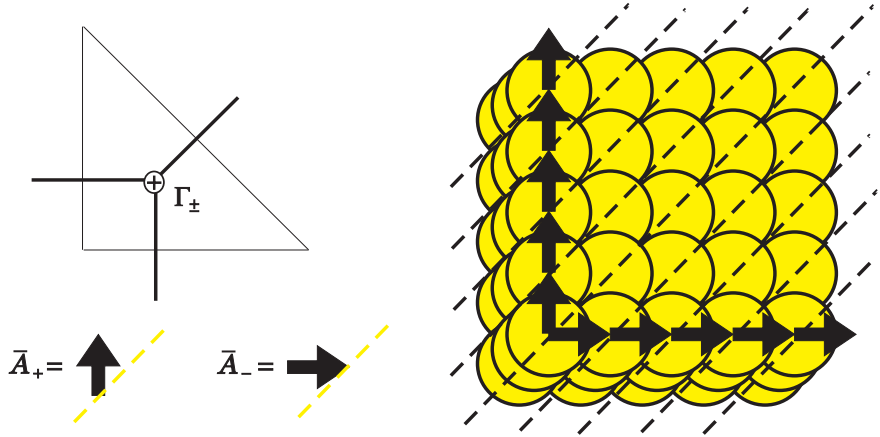} 
\begin{quote}
\caption{\emph{Toric diagram for $\mathbb{C}^3$ (left) consists of one $\oplus$ vertex. Operators $\overline{A}_{\pm}$  involve a single $\G_{\pm}$, and have a simple arrow (lower left), as follows from figure \ref{fig-arrows}. The correlator (\ref{C3-crystal}) is translated into a sequence of arrows, with rotated dashed lines representing insertions of interlacing two-dimensional partitions. The resulting figure (right) represents plane partitions crystal model, the same as in figure \ref{fig-C3}, but now seen from the bottom. 
}} \label{fig-C3new}
\end{quote}
\end{center}
\end{figure}

Let us reconsider $\mathbb{C}^3$ geometry which motivated our discussion in section \ref{ssec-C3example}. 
In this case, the dual toric diagram consists just of one triangle, see fig. \ref{fig-C3new} (left), so there is just one vertex and only one color $\widehat{Q}_0\equiv \widehat{Q}$, and the operators (\ref{Apm}) take form
$$
\overline{A}_{\pm} = \G_{\pm}(1) \widehat{Q}.
$$
In consequence the BPS partition function (\ref{Z-Omega}) takes exactly the form (\ref{C3-crystal}).

The crystal structure can be read off from a sequence of arrows associated to $\widehat{A}_{\pm}$ operators, following the rules in figure \ref{fig-arrows}. This gives rise to the crystal shown in fig. \ref{fig-C3new} (right). This is the same crystal as in fig. \ref{fig-C3}, which represents plane partitions, however now seen from the opposite side. 

\subsubsection{Orbifolds $\mathbb{C}^3 / \mathbb{Z}^{N+1}$}    \label{ssec-OrbiC3}

\begin{figure}[htb]
\begin{center}
\includegraphics[width=\textwidth]{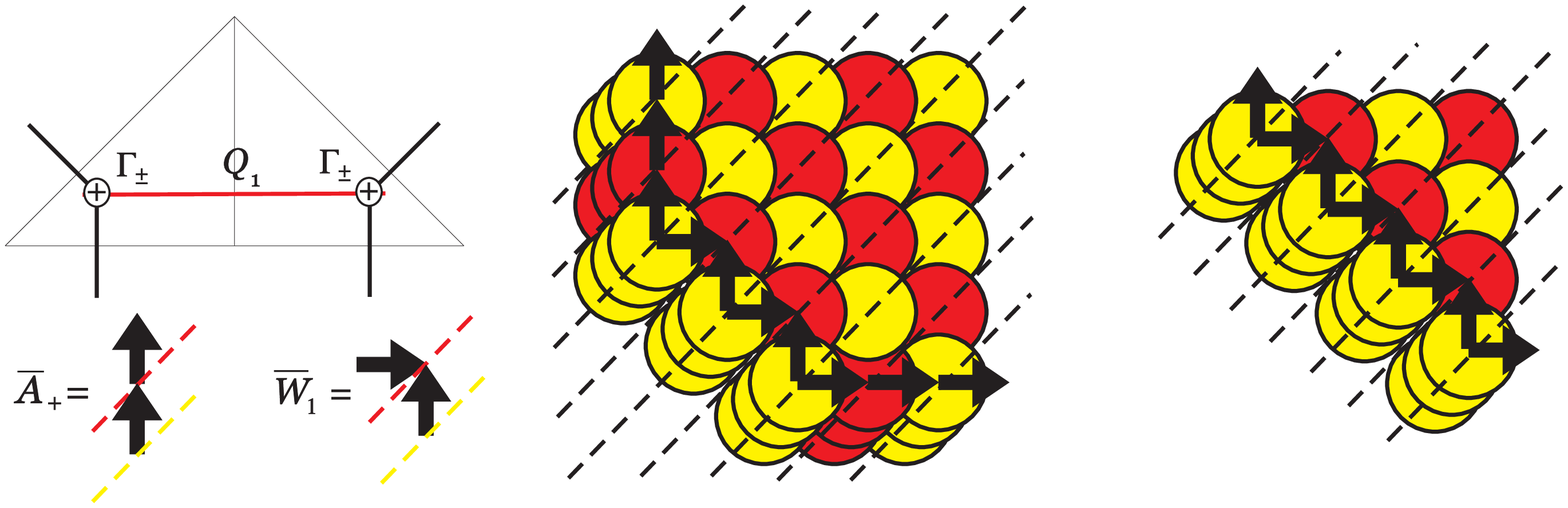} 
\begin{quote}
\caption{\emph{Toric diagram for the resolution of $\mathbb{C}^3/ \mathbb{Z}^{N+1}$ geometry has $N+1$ vertices of the same type $\oplus$. Left: toric diagram for $N=1$ and arrow representation of $\overline{A}_+$ and $\overline{W}_1$. In the non-commutative chamber this leads to the same plane partition crystal as in fig. \ref{fig-C3new}, however colored now in yellow and red. Middle: for the chamber with positive $R$ and $2<B<3$ the crystal develops two additional corners and its generating function reads $Z_{2|1} = \langle \Omega_+|(\overline{W}_1)^2|\Omega_-\rangle$. Right: for negative $R$ and positive $n-1<B<n$ the crystal is finite along two axes 
and develops $n-1$ yellow corners; its generating function for the case of $n=5$ shown in the picture reads $\widetilde{Z}_{5|1} = \langle 0 |(\overline{W}_1)^5|0\rangle$ (two external arrows, corresponding to $\G_-$ acting on $\langle 0|$ and $\G_+$ acting on $|0\rangle$, are suppressed.)}} \label{fig-OrbiC3}
\end{quote}
\end{center}
\end{figure}

Now we consider the resolution of $\mathbb{C}^3 / \mathbb{Z}^{N+1}$ orbifold. In this case the toric diagram takes form of a triangle of area $(N+1)/2$, see fig. \ref{fig-OrbiC3} (left). There are $N$ independent $\mathbb{P}^1$'s and $N+1$ vertices of the same $t_i=+1$, and operators in (\ref{Apm}) take the form
$$
\overline{A}_{\pm} = \G_{\pm}(1) \widehat{Q}_1 \G_{\pm}(1) \widehat{Q}_2 \ldots \G_{\pm}(1) \widehat{Q}_{N} \G_{\pm}(1) \widehat{Q}_0.
$$
In the non-commutative chamber the corresponding crystal consists of plane partitions, however with slices colored periodically in $N+1$ colors. The partition function in the non-commutative chamber is given by (\ref{Z-Omega}). 

If we turn on an arbitrary $B$-field through a fixed $\mathbb{P}^1$, the structure of wall-crossing operators gives rise to modified containers, see e.g. fig. \ref{fig-OrbiC3} (middle). In particular enlarging the $B$-field by one unit adds one more yellow corner to the crystal. 

The crystals corresponding to $R<0$ are also easy to find. In the extreme chamber we get a trivial (empty) crystal, representing a single D6-brane (\ref{Ztilde-1}). Adding wall-crossing operators results in a crystal with several corners, finite along two axis (and extending infinitely along the third axis), as shown in fig. \ref{fig-OrbiC3} (right).


\subsubsection{Resolved conifold}    \label{ssec-ExampleConifold}

\begin{figure}[htb]
\begin{center}
\includegraphics[width=0.9\textwidth]{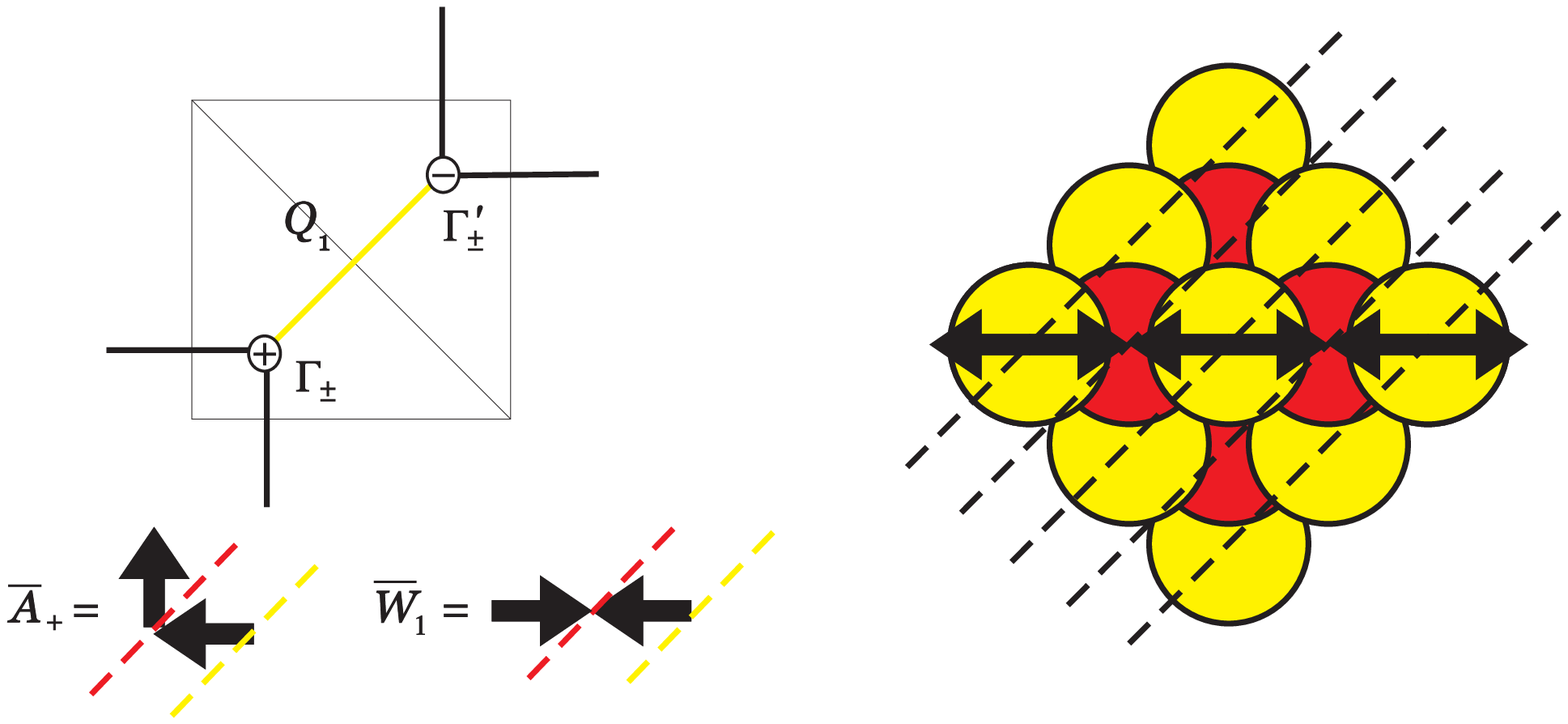} 
\begin{quote}
\caption{\emph{Left: toric diagram for the conifold and arrow representation of $\overline{A}_+$ and $\overline{W}_1$. Right: for chambers with negative $R$ and positive $n-1<B<n$ the crystals are given by finite pyramid partitions with $n-1$ additional corners, represented by $n-1$ stones in the top row (the figure shows the case $n=4$). The generating function is given by $\widetilde{Z}_{n|1} = \langle 0|(\overline{W}_1)^n|0\rangle$ which reproduces the result (\ref{ZZn}). 
}} \label{fig-pyramid-coni-inf3}
\end{quote}
\end{center}
\end{figure}

We already presented pyramid crystals for the conifold in section \ref{ssec-crystals}. They arise from our formalism as follows.
The dual toric diagram for the conifold, see fig. \ref{fig-pyramid-coni-inf3} (left), consists of two triangles and encodes a single ($N=1$) $\mathbb{P}^1$. Two vertices of the toric diagram correspond to two colors $\widehat{Q}_1$ and $\widehat{Q}_0$, so that
$$
\widehat{Q} = \widehat{Q}_1 \widehat{Q}_0, \qquad \qquad q=q_1 q_0.
$$
The operators (\ref{Apm}) in this case read
$$
\overline{A}_{\pm}(x) = \G_{\pm}(x) Q_1 \G'_{\pm}(x) Q_0,
$$
while (\ref{Aplus}) and (\ref{Aminus}) are
$$
A_+(x) = 
\G_+(xq ) \G'_+(x q / q_1), \qquad \qquad A_-(x) = 
\G_-(x) \G'_-(xq_1),
$$
and they satisfy
$$
A_+(x) A_-(y) = \frac{(1+xy q / q_1)(1+xy q q_1)}{(1-xy q)^2}A_-(y) A_+(x).
$$

The quantum states (\ref{Omega-plus}) and (\ref{Omega-minus}) take form
\bea
|\Omega_-\rangle & = & 
A_-(1) A_-(q) A_-(q^2)\ldots |0\rangle,  \\
\langle \Omega_+ | & = & 
\langle 0| \ldots A_+(q^2) A_+(q) A_+(1) \\
\eea
and the wall-insertion operators (\ref{Wp}) (\ref{Wpprim}) are
\be
W_1(x) = \G_-(x) Q_1 \G'_+(x) Q_0, \qquad \qquad   W'_1(x) = \G_+(x) Q_1 \G'_-(x) Q_0.     \label{Wp-conifold}
\ee

\begin{figure}[htb]
\begin{center}
\includegraphics[width=0.8\textwidth]{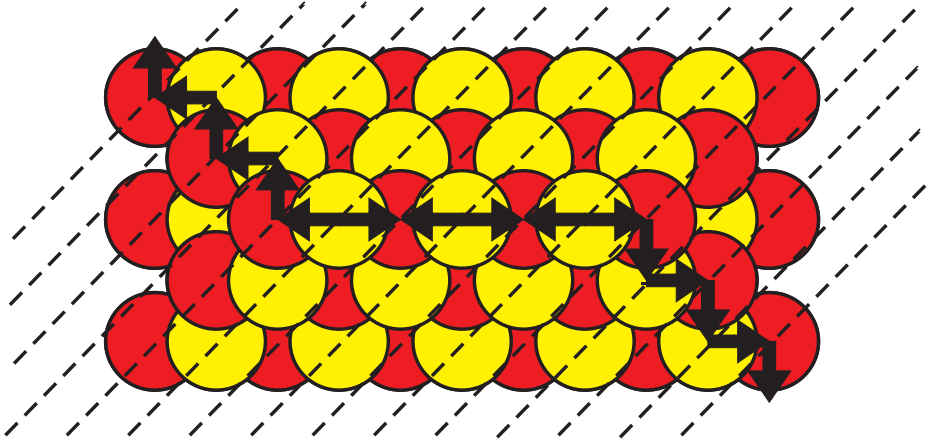} 
\begin{quote}
\caption{\emph{Conifold crystal in the chamber with positive $R$ and $2<B<3$ takes form of pyramid partitions with $3$ stones in the top row. Its generating function is given by $Z_{2|1} = \langle \Omega_+|(\overline{W}_1)^2|\Omega_-\rangle$.}} \label{fig-conifold-crystal}
\end{quote}
\end{center}
\end{figure}

Therefore the fermionic correlators take form
\bea
Z_{n|1} & = & \langle \Omega_+ |(\overline{W}_1)^n| \Omega_- \rangle,    \\
\widetilde{Z}_{n|1} & = & \langle 0|(\overline{W}_1)^{n} |0 \rangle.   \label{coni-finite-crystal}
\eea
and encode generating functions (\ref{Zn}) and (\ref{ZZn}) introduced in section \ref{ssec-crystals}. 
In the non-commutative chamber we get the result found first in \cite{Szendroi},  $Z_{0|1}=\langle\Omega_+|\Omega_-\rangle$, while a single D6-brane is encoded in $\widetilde{Z}=\langle 0|0\rangle =1$. These crystals are shown in 
 fig. \ref{fig-pyramid-coni-inf3} (right) and \ref{fig-conifold-crystal}.

\section{Matrix models and open BPS generating functions}    \label{sec-matrix}

In this section we explain how matrix model formalism can be applied to analyze BPS counting functions. In the first part, \ref{ssec-matrixFermion}, we explain how to relate fermionic formalism, derived in the previous section, to matrix model representation. In section \ref{ssec-matricNCDT}, we illustrate how to construct matrix models for the closed non-commutative chamber. In subsection \ref{ssec-matrix-conifold} we analyze in detail BPS generating functions for the conifold for all chambers with $R>0$, and derive corresponding spectral curves. We discuss how these curves relate to (and generalize) mirror curves, which we find (as we should) in the commutative chamber. In subsection \ref{ssec-matrixOpen} we reveal that matrix model representation in fact encodes open BPS generating functions, which can be identified with matrix model integrands.


\subsection{Matrix models from free fermions}    \label{ssec-matrixFermion}

Let us explain how to relate fermionic representation of BPS amplitudes, introduced in section \ref{ssec-triangulate}, to matrix models. This relies on introducing into fermionic correlators representing BPS generating functions, such as (\ref{Z-Omega}) or (\ref{ZRposBpos-Results}), a special representation of the identity operator $\mathbb{I}$. The representation we are interested in also consists of infinite product of vertex operators and arises as follows \cite{OSY}. Firstly, we can use the representation as a complete set of states $\mathbb{I} = |R\rangle\langle R|$, which represent two-dimensional partitions. Using orthogonality relations of $U(\infty)$ characters $\chi_R$, and the fact that these characters are given in terms of Schur functions $\chi_R=s_R(\vec{z})$ for $\vec{z}=(z_1,z_2,z_3,\ldots)$, we can write
\bea
\mathbb{I} & = & \sum_R |R\rangle\langle R|  = \sum_{P,R} \delta_{P^t R^t} |P\rangle\langle R| = \nonumber \\
& = & \int \mathcal{D}U \sum_{P,R} s_{P^t}(\vec{z}) \overline{s_{R^t} (\vec{z})} |P\rangle\langle R| = \nonumber \\
& = & \int \mathcal{D}U \Big(\prod_{\alpha} \G_-'(z_{\alpha})|0\rangle  \Big)  \Big(\langle 0 | \prod_{\alpha} \G_+'(z^{-1}_{\alpha}) \Big).    \label{identity}
\eea
When such a representation of the identity operator is introduced into (\ref{Z-Omega}) or (\ref{ZRposBpos-Results}) (or any other correlator of similar structure) we can commute away $\G_{\pm}^{t_i}$ operators and get rid of operator expressions. For example, inserting the above identity operator in the string of $\overline{A}_+$ operators in (\ref{Z-Omega-Wn}),
leads to a matrix model with the unitary measure
\bea
Z_n & = & \langle 0 | \prod_{i=k}^{\infty}  \overline{A}_+(1) | \mathbb{I} | \prod_{j=0}^{k-1}  \overline{A}_+(1) | \overline{W}^n  |\Omega_-\rangle = \nonumber \\
& = & \int \mathcal{D}U  \ \langle 0 | \prod_{i=k+1}^{\infty}  \overline{A}_+(1) |
\prod_{\alpha} \G_-'(z_{\alpha})|0\rangle  \langle 0 | \prod_{\alpha} \G_+'(z^{-1}_{\alpha}) | \prod_{j=0}^{k}  \overline{A}_+(1) | \overline{W}^n | \Omega_-\rangle =   \nonumber   \\
& = & f^k_{n}(q,Q_i) \int \mathcal{D}U \prod_{\alpha} e^{-\frac{1}{g_s} V^{k}_{n}(z_{\alpha})}.  \label{Zkn-int}
\eea
The product over $\alpha$ represents distinct eigenvalues $z_{\alpha}$. Note that we have inserted $\mathbb{I}$ at the position $k$ in the string of $\overline{A}_+(1)$ operators. In particular this affects the form of the resulting potential $V^k_n(z)$. Moreover, apart from matrix integral, we find some overall factors $f^k_n(q,Q_i)$ which take form of various infinite products. They arise, in a generic chamber, from commutations between $\G_{\pm}$ ingredients of wall-crossing operators, and $\G_{\mp}$ ingredients of $|\Omega_{\mp}\rangle$ states. In the closed non-commutative chamber $n=0$ these factors are trivial, $f_{n=0}(q,Q_i)=1$, and they largely simplify in the commutative chamber $n\to \infty$.

\begin{figure}[htb]
\begin{center}
\includegraphics[width=0.4\textwidth]{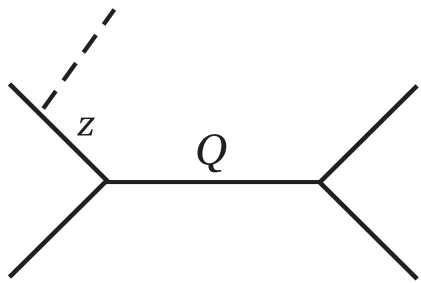} 
\begin{quote}
\caption{\emph{Brane associated to the external leg of a toric diagram (of a conifold in this particular case). Closed string parameter is denoted by $Q$ and open string parameter by $z$.}} \label{fig-conibrane}
\end{quote}
\end{center}
\end{figure}

There is a large freedom in choosing the value of $k$, and it is natural to ask if this choice has some physical interpretation. It was argued in \cite{openWalls} that this is indeed the case, and the choice of $k$ is equivalent to the choice of open BPS chamber (open BPS chambers were introduced in section \ref{ssec-M}). In particular, it turns out that the open generating parameter can be identified with matrix eigenvalues $z_{\alpha}$, and the open BPS generating function (\ref{Zbps-Ztop2}) in the open chamber labeled by $k$ can be identified with matrix integrand
\be
\cZ^{open}_{BPS} = 
e^{-\frac{1}{g_s} V^{k}_{n}(z)}.   \label{claim}
\ee
Even though the overall factors $f^k_n$ in (\ref{Zkn-int}) may involve closed moduli $Q_i$, they do not involve open moduli $z$. In this sense the matrix integrand is well defined, and up to some simple identification can be identified with open BPS generating function. 
This identification of parameters amounts to the shift $z\to-zq^{1/2}$ (to match earlier M-theory convention with half-integer powers of $q$, to integer powers of $q$ in the fermionic formalism), as well as identification of K{\"a}hler parameters considered in M-theory derivation with parameters $\mu_i$ introduced below. We also note that the BPS generating function in (\ref{Zbps-Ztop2}) is determined by the open topological string partition function associated to the external axis of the toric diagram, as in figure \ref{fig-conibrane}. As we will also see, the value of the above integral (\ref{Zkn-int}) can be related to some more general Calabi-Yau geometry $Y$. 

\subsection{Matrix models for the non-commutative chamber}   \label{ssec-matricNCDT}

In this section we illustrate the relation between BPS counting and matrix models in case of the non-commutative chamber $n=0$, and the choice of open chamber also $k=0$. This corresponds to the insertion of the identity representation (\ref{identity}) exactly in between $|\Omega_{\pm}\rangle$ states in (\ref{Z-Omega}). In this $n=0$ case no factor $f^k_n$ in (\ref{Zkn-int}) arises, and we obtain matrix models with potentials which can be expressed in terms of the following version of the theta function
$$
\Theta(z;q) = \prod_{j=0}^{\infty} (1 + zq^j)(1+q^{j+1}/z).
$$
For a general geometry of the form shown in fig. \ref{fig-strip}, with types of vertices given by $t_i$, corresponding matrix models take form
\be
Z = \int dU \prod_{\alpha} e^{-\frac{1}{g_s} V(z_{\alpha})} = \int dU \prod_{\alpha} \prod_{l=0}^N \Theta\big(t_{l+1} z_{\alpha} (q_1\cdots q_l) ; q\big)^{t_{l+1}},   \label{Zn0k0}
\ee
where integral is over unitary matrices of infinite size, $N=\infty$. Special cases of this result include:
\begin{itemize}
\item for $\mathbb{C}^3$ the result (\ref{Zn0k0}) provides a matrix model representation of MacMahon function $Z=M(1)=\prod_{k=1}^{\infty}(1-q^k)^{-k}$ in terms of a matrix model of the form (\ref{Zn0k0}) with the integrand
$$
e^{-\frac{1}{g_s} V(z)}  =  \prod_{j=0}^{\infty} (1 + zq^j)(1 + q^{j+1}/z) = \Theta(z;q).
$$
\item for the conifold we obtain a representation of the pyramid partition generating function (\ref{ZnDT}) (with $n=0$) in terms of a matrix model with the integrand
$$
e^{-\frac{1}{g_s} V(z)}  =  \prod_{j=0}^{\infty} \frac{(1 + zq^j)(1 + q^{j+1}/z)}{(1 + Qzq^j)(1 + \frac{q^{j+1}}{Qz})} = \frac{\Theta(z;q)}{\Theta(Qz;q)}
$$
\item for $\mathbb{C}^3/\mathbb{Z}^{N+1}$ we have $t_p=+1$ for all $p$ and we find matrix model representation of the BPS generating function in terms of a matrix model with the integrand
\bea
e^{-\frac{1}{g_s} V(z)} &  = & \prod_{j=0}^{\infty} (1 + zq^j)(1 + q^{j+1}/z) \cdots (1 + (q_1\cdots q_N) zq^j)(1 + \frac{q^{j+1}}{(q_1\cdots q_N) z}) = \nonumber \\
& = & \prod_{l=0}^N \Theta((q_1\cdots q_l)z;q)  \nonumber
\eea
\end{itemize}


\subsection{Matrix model for the conifold -- analysis}   \label{ssec-matrix-conifold}

In this section we illustrate how matrix model techniques can be used in the context of models which arise for BPS counting. We focus on the conifold matrix model in arbitrary closed BPS chamber $n$, and fixed $k=0$. In this case the result (\ref{Zkn-int}) takes form
(after the redefinition $Q=-q_1 q^n$) 
\bea
Z_n & = & M(q)^2 \prod_{j=1}^{\infty} (1-Q q^j)^j (1-Q^{-1}q^{j+n})^{j+n} = \label{MM-Zn-conifold} \\
& = & f_n(q,Q) \int dU \prod_{\alpha} \prod_{j=0}^{\infty}  \frac{ (1+z_{\alpha} q^{j+1})\ (1+q^{j}/z_{\alpha})}{ (1+z_{\alpha} q^{j+n+1}/Q)\ (1+q^j Q/z_{\alpha})} = f_n(q,Q) Z_{matrix},    \nonumber
\eea
with
\be
f_n(q,Q) = M(q) \frac{\prod_{j=1}^{\infty} (1-q^{n+j}/Q)^n}{M(q^n,q)},       \label{fn-conifold}
\ee
with MacMahon function $M(q)$ defined in (\ref{macmahon}), and with the following generalized MacMahon function
\be
M(z,q) = \prod_{i=1}^{\infty} \frac{1}{(1 - z q^i)^i}  \label{macmahon-zq}
\ee
In particular, in the non-commutative chamber $f^{conifold}_0=1$, and in the commutative chamber $f^{conifold}_{n\to\infty} = M(q)$ which represents topological string degree zero contributions. The result (\ref{MM-Zn-conifold}) implies that the value of the matrix model integral (without the prefactor $f_n$) is equal to 
\bea
Z_{matrix} & = & \frac{Z_n}{f_n(q,Q)} = 
M(q) \prod_{j=1}^{\infty} \frac{(1-Qq^j)^j(1-\mu q^j)^j}{(1-\mu Q q^j)^j } = \nonumber \\
& = & \int dU \prod_{\alpha} \prod_{j=0}^{\infty}  \frac{ (1+z_{\alpha} q^{j+1})\ (1+q^{j}/z_{\alpha})}{ (1+z_{\alpha} q^{j+n+1}/Q)\ (1+q^j Q/z_{\alpha})},   \label{prod-int-coni}
\eea
where $\mu=q^n/Q$. 

Now we wish to analyze the matrix model $Z_{matrix}$. We parametrize the 't Hooft coupling and the chamber dependence respectively by 
\be
T=g_s N,\qquad \qquad \tau = n g_s.
\ee
As our models correspond to $U(\infty)$ matrices, ultimately we are interested in the limit
\be
T\to\infty,\qquad \quad g_s=const, \qquad \quad Q=const,   \label{limit}
\ee
for each fixed chamber (i.e. fixed $n$ and therefore $\tau$).  The non-commutative chamber corresponds to $\tau=0$, while $\tau\to\infty$ represents the topological string chamber.

Using the expansion of the quantum dilogarithm
\be
\log \prod_{i=1}^{\infty} \big(1 - z q^i\big) = -\frac{1}{g_s} \sum_{m=0}^{\infty} \Li_{2-m}\big(z\big) \frac{B_m g_s^m }{m!} , \label{qdilog-expand}
\ee
and the redefinition of the unitary measure (\ref{measure-T}) we find, to the leading order in $g_s$, the following matrix model potential
\be
V_{\tau} = T\log(z) + Li_2(-z) + \Li_2\big(-\frac{1}{z}\big) - \Li_2\big(-\frac{Q}{z}\big) - \Li_2\big(-\frac{z}{Q e^{\tau}}\big), \label{Vconifold}
\ee
so that 
\be
\partial_z V_{\tau} = \frac{T - \log(z+Q) +\log\big(1+\frac{z}{Qe^{\tau}}\big)}{z},     \label{partialV-conifold}
\ee

Now we wish to solve the model (\ref{MM-Zn-conifold}) in the small $g_s$ limit. Firstly we need to find the resolvent $\omega(p)$, which can be done using the Migdal integral (\ref{Migdal}), and careful derivation is presented in \cite{OSY}. As we expect one-cut solution of our model, from the Migdal integral we get an expression in terms of the end-points of this cut $a$ and $b$. The normalization condition (\ref{Migdal-norm}) imposes two constraints, for terms of order $p^0$ and $p^{-1}$ in the resolvent, which take form
\bea
\frac{\sqrt{a+Q}-\sqrt{b+Q}}{\sqrt{a+Qe^{\tau}}-\sqrt{b+Qe^{\tau}}} & = &  Q^{1/2} e^{(\tau+T)/2}      \label{ab-eq1}  \\
\frac{\sqrt{(a+Q)b}-\sqrt{(b+Q)a}}{\sqrt{(a+Qe^{\tau})b}-\sqrt{(b+Qe^{\tau})a}} & = & Q^{1/2}  e^{-(\tau+T)/2}     \label{ab-eq2}
\eea
These constraints can be solved in the exact form, with result
\bea
a & = & -1 + \epsilon^2 \frac{(1-\mu)(1-\mu\epsilon^2) + (1-Q)(1+\mu\epsilon^2 - 2\mu)}{(1-\mu\epsilon^2)^2} + \label{aAll} \\
& & + 2 i \epsilon \frac{\sqrt{(1-Q)(1-\epsilon^2)(1-\mu)(1-Q\mu\epsilon^2)}}{(1-\mu\epsilon^2)^2},  \nonumber  \\
b & = & -1 + \epsilon^2 \frac{(1-\mu)(1-\mu\epsilon^2) + (1-Q)(1+\mu\epsilon^2 - 2\mu)}{(1-\mu\epsilon^2)^2} + \label{bAll} \\
& & - 2 i \epsilon \frac{\sqrt{(1-Q)(1-\epsilon^2)(1-\mu)(1-Q\mu\epsilon^2)}}{(1-\mu\epsilon^2)^2} ,  \nonumber
\eea
where we introduced
\be
\epsilon = e^{-T/2},\qquad \qquad  \mu = \frac{1}{Qe^{\tau}}.
\ee
Substituting these end-points back to the formula for the resolvent we find
\be
\omega_{\pm}(p) = \frac{1}{pT} \log \Big(\frac{(1+\mu\epsilon^2)p + (1+Q\epsilon^2) \mp (1-\mu\epsilon^2) \sqrt{(p-a)(p-b)}}{2 e^{-T} (p+Q)} \Big).    \label{omega-coni-muSmall}
\ee
As a check, this result indeed satisfies the consistency condition (\ref{omegaPM})
$$
\omega_+(p) + \omega_-(p) =\frac{\partial_pV_{\tau}(p)}{T},       
$$
with $V_{\tau}$ given in (\ref{Vconifold}). From the knowledge of the resolvent we can also determine eigenvalue density along the cut (\ref{rho-omega})
$$
\rho(p) = \frac{1}{pT} \log\left( \frac{ (1+\mu\epsilon^2)p + 1 +Q\epsilon^2 - (1-\mu \epsilon^2) \sqrt{(p-a)(p-b)}}{ (1+\mu\epsilon^2)p + 1+Q\epsilon^2 + (1-\mu \epsilon^2)\sqrt{(p-a)(p-b)}}  \right),
$$
as well as the spectral curve. Writing $x=pT\omega(p)$ and $p=e^y$, and after a few simple rescalings we find that the spectral curve takes form
\be
  e^{x+y} + e^x + e^y + Q_1\ e^{2x} + Q_2\ e^{2y} + Q_3= 0,
\label{symmetricresolution}
\ee   
where
\begin{align}
\begin{split}
& Q_1 = \epsilon^2 \, 
\frac{1+\mu Q}{(1+\mu \epsilon^2)(1+Q\epsilon^2)}, \\
& Q_2 = \mu \,
 \frac{1+Q\epsilon^2}{(1+\mu Q)(1+\mu\epsilon^2 )}, \\
& Q_3 = Q \, \frac{1+\mu \epsilon^2}{(1+\epsilon^2 Q)(1+\mu Q)}.
\end{split}
\end{align}

\begin{figure}[htb]
\begin{center}
\includegraphics[width=0.5\textwidth]{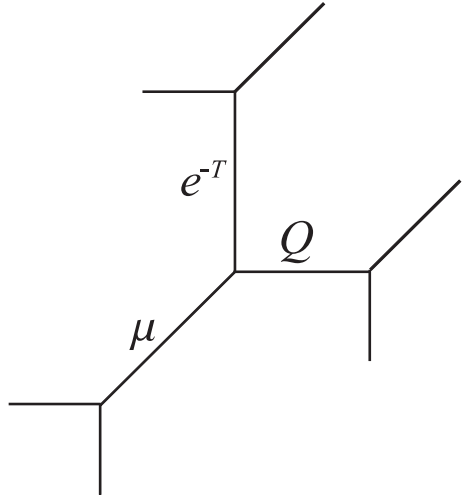} 
\begin{quote}
\caption{\emph{The spectral curve for the conifold matrix model (\ref{MM-Zn-conifold}) in arbitrary closed BPS chamber coincides with the mirror curve of the closed topological vertex geometry, whose toric diagram is shown above. This geometry has three $\mathbb{P}^1$ moduli, which encode conifold size $Q$, the closed BPS chamber via $\mu$, and finite 't Hooft coupling via $e^{-T}$.}} \label{fig-closedvertex}
\end{quote}
\end{center}
\end{figure}

The above curve is given by a symmetric function of $Q$, $\mu = Q^{-1}q^n$ and $\epsilon^2 
= e^{-T}$. Apparently this is a mirror curve of the so-called \emph{closed topological vertex} geometry, which is a Calabi-Yau manifold arising from a symmetric resolution of $\mathbb{C}^3/ \mathbb{Z}_2 \times \mathbb{Z}_2$ orbifold, see fig. \ref{fig-closedvertex}. This geometry has three moduli, i.e. the original K{\"a}hler moduli $Q$ of the resolved conifold, the chamber parameter $n$ (encoded in $\mu$) and the 't Hooft parameter $T$, which are all unified in a geometric way in our matrix model. Moreover, the fractional coefficients in the curve equation (\ref{symmetricresolution}) encode the correct mirror map for the closed topological vertex geometry (and to the linear order, these coefficients are just $Q$, $\mu$ and $e^{-T}$).

In the BPS counting problem we are interested in, as follows from the form of the identity operator (\ref{identity}), ultimately we need to consider matrices of infinite size. We also need to keep fixed $g_s$, so we should consider the limit of $T \to \infty$, or equivalently $\epsilon \to 0$. Up to a linear shift of $x$ and $y$, the equation (\ref{symmetricresolution}) in this limit becomes
\be
  \mu\ e^{2y} + e^{x+y} + e^x + (1 + Q\mu)\ e^y + Q = 0.
\label{SPPcurve}
\ee
The manifold corresponding to this curve is the \emph{suspended pinch point} (SPP) geometry,
with $Q$ and $\mu$ encoding flat coordinates representing sizes of its two $\mathbb{P}^1$'s. Having found the SPP mirror curve, let us also make the following remarks.

Firstly, we see that not only the spectral and mirror curves agree, but moreover the matrix integral (\ref{MM-Zn-conifold}) reproduces (after the identiciation $q=q_s$)  the full topological string partition function of the SPP geometry at finite $g_s$, which is known to take form
\be
 \cZ^{\rm SPP}_{{\rm top}}(q_s, Q, \mu) 
= \prod_{k=1}^\infty \frac{(1-Qq_s^k)^k(1-\mu q_s^k)^k}
{(1-q_s^k)^{3k/2}(1 - \mu Q q_s^k)^k}.     \label{ZtopSPP}
\ee
for K{\"a}hler parameters $Q$ and $\mu$. This confirms that our result makes sense, although this also means that the terms in lowest order in $g_s$ in the potential reproduce the full $g_s$ dependence of the partition function. It would be nice to prove rigorously that higher $g_s$ corrections to the potential indeed do not affect the total partition function. This appears to be a very special feature of matrix models integrands which can be expressed -- as is the case for (\ref{prod-int-coni}) -- in terms of infinite products of the form $\prod_k (1 - xq^k)$. One proof of such statement in a related situation (although in addition taking advantage of a special phenomenon of the \emph{arctic circle}) can be found in \cite{eynard-planch}.

Secondly, it is natural to conjecture that the total partition function of the matrix model, for finite $T$, should reproduce (at least up to some MacMahon factor) the topological string partition function for the closed topological vertex which reads
$$
 Z^{\rm total}_{{\rm matrix}}(q, Q, \mu,\epsilon^2) 
= \prod_{k=1}^\infty (1-q^k)^k \cdot \prod_{k=1}^\infty \frac{(1-Qq^k)^k(1-\mu q^k)^k (1-\epsilon^2 q^k)^k(1-Q\mu \epsilon^2 q^k)^k}
{(1 -  Q\mu q^k)^k (1 - \mu \epsilon^2 q^k)^k (1 - Q\epsilon^2 q^k)^k}.    
$$

Finally,  we also note that in the limit $Q,\mu\to 0$ our model reduces to the 
Chern-Simons matrix model discussed in \cite{Marino,lens-matrix}. Indeed, in this limit the potential (\ref{Vconifold}) reproduces gaussian potential \cite{lens-matrix}
$$
V_{Q\to 0,n\to\infty} = -\frac{1}{2}(\log z)^2 = -\frac{1}{2}u^2.
$$
In this case the resolvent (\ref{omega-coni-muSmall}) reduces to 
$$
\omega^{conifold}_{\pm}(p)_{\mu=Q=0} = \frac{1}{pT} \log \Big(\frac{p + 1 \mp \sqrt{(p+1)^2- 4p e^{-T}}}{2p e^{-T}} \Big) ,
$$
and agrees\footnote{Instead of introducing $T\log z$ term to the potential to get the standard Vandermonde determinant, the solution in \cite{Marino} involves completing the square, which leads to a redefinition $p_{here}=p_{\textrm{\cite{Marino}}}e^T$. Due to a different sign of $g_s$ we also need to identify 't Hooft couplings as $T_{here}=-t_{\textrm{\cite{Marino}}}$. Taking this rescaling into account, our cut endpoints (\ref{aAll},\ref{bAll}) with $Q=\mu=0$ also coincide with those in \cite{Marino}.} with the resolvent of the old Chern-Simons matrix model found in \cite{Marino,lens-matrix}, which is known to reproduce the Chern-Simons partition function
\be
Z_{Chern-Simons} = \frac{M(q_s)}{M(e^{-T},q_s)},  \label{ZCS}
\ee
with 't Hooft coupling $T$ identified with the K{\"a}hler parameter of the conifold. A manifestation of this result is seen directly in the product formula in (\ref{prod-int-coni}), which is an expression for matrices of infinite size (i.e. $T\to\infty$): for $Q,\mu\to 0$ the matrix model partition function reduces to the MacMahon function $M(q)$, which is consistent with (\ref{ZCS}) in $T\to \infty$ limit. For finite $T$, also the spectral curve of the matrix model with the above gaussian potential reproduces the conifold mirror curve (\ref{mirrorcurves}) of the size given by the 't Hooft coupling
$$
x + p + xp + x^2 e^{-T}  = 0. 
$$
Let us also mention that the MacMahon function arising in $T\to\infty$ limit of this model, and in fact the entire Chern-Simons partition function, can be obtained -- following the postulate of the remodeling conjecture \cite{BKMP} -- from the topological recursion applied to the above curve \cite{vb-ps}. It turns out that the topological recursion associates a square root of the MacMahon function to each patch of $\mathbb{C}^3$, in agreement with (\ref{MacMahon-C3}), and building the toric Calabi-Yau three-fold by glueing such patches is mirrored by the pant decomposition of the mirror curve. However, the behavior of such constant contributions to the partition function in matrix models is subtle, and in general may not agree with the topological string (or topological recursion) result, even if a spectral curve of a matrix model reproduces an appropriate mirror curve. For example, matrix models constructed in \cite{eynard-planch,SW-matrix} reproduce correct mirror curves (in particular, for the conifold) and encode correct non-constant contributions to the partition function, however by construction do not involve any factor of MacMahon function.


\subsection{Matrix models and open BPS generating functions}   \label{ssec-matrixOpen}

In this section we finally consider arbitrary closed and open BPS chamber, so that matrix models take most general form. Analyzing the case of $\C^3$, conifold and $\C^3/\Z_2$ we illustrate the claim (\ref{claim}) that open BPS generating functions can be identified with integrand of matrix models. Also for this reason our analysis is only on the level of these integrands, however it would also be interesting to understand the corresponding spectral curves. 


\subsubsection{$\C^3$} 

We recall that the open topological string amplitude for a brane in $\mathbb{C}^3$ is given by the quantum dilogarithm (\ref{qdilog}). The condition for the central charge (\ref{Zopen}) and the general formula (\ref{Zbps-Ztop2}) imply that in the open chamber labeled by $k$ the open BPS generating function reads
\be
\cZ^{open}_{k} = \prod_{i=1}^{\infty} (1 - z q_s^{i-1/2}) \prod_{j>k}^{\infty} (1 - z^{-1} q_s^{j-1/2}).    \label{ZbpsC3-1}
\ee
In one extreme chamber $k=0$ the generating function
is equal (up to the overall $q^{1/24}$) to a theta function, and so it is a modular form, as explained in section \ref{ssec-M}. On the other hand, for $k\to\infty$, the generating function  $Z_{k\to\infty}$ reduces to the open topological string amplitude (\ref{Ztop-open}).

\begin{figure}[htb]
\begin{center}
\includegraphics[width=0.35\textwidth]{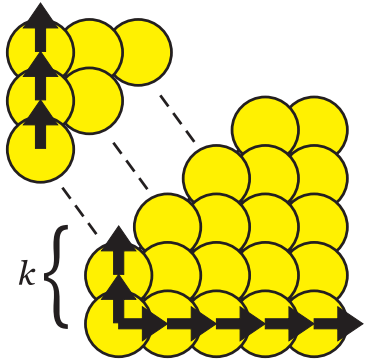} 
\begin{quote}
\caption{\emph{Factorization of $\mathbb{C}^3$ crystal which leads to open BPS generating functions. The size $k$ encodes the open BPS chamber.}} \label{fig-C3split2}
\end{quote}
\end{center}
\end{figure}

Now we present how the result (\ref{ZbpsC3-1}) arises from the matrix model viewpoint. To start with, we again consider fermionic representation. Following results of section \ref{ssec-C3}, in this case $\overline{A}_+(1) = \G_+(1) \widehat{Q}$ and to the geometry of $\mathbb{C}^3$ we associate a state
$$
|\Omega_-\rangle = \prod_{i=1}^{\infty} \G_-(q^i)|0\rangle,
$$
and similarly for $\langle\Omega_+|$. There is a single closed string chamber in which the generating function $Z=\langle\Omega_+|\Omega_-\rangle = M(q)$ is given by the MacMahon function.
Following the prescription (\ref{Zkn-int}) we insert the operator $\mathbb{I}$ at the location $k$. This gives
\bea
Z & = & M(q) = \langle 0 | \prod_{i=k}^{\infty} \overline{A}_+(1) | \mathbb{I} |  \prod_{j=0}^{k-1} \overline{A}_+(1) | \Omega_-\rangle = \nonumber  \\
& = & f^k(q) Z_{matrix},    \label{C3open-matrix}
\eea 
where the matrix integral is given by
$$
Z_{matrix} = \int \mathcal{D}U \prod_{\alpha} \prod_{j=1}^{\infty} (1 + z_{\alpha}q^j) \prod_{i=k}^{\infty} (1 + z_{\alpha}^{-1}q^j) ,
$$
and
$$
f^k(q) = \prod_{i=1}^{k} \prod_{j=0}^{\infty} \frac{1}{1-q^{i+j}} = \frac{M(1)}{M(q^k,q)},
$$
with the generalized MacMahon function $M(q^k,q)$ defined in (\ref{macmahon-zq}). Matrix model integrand in $Z_{matrix}$ indeed reproduces open BPS generating function (\ref{ZbpsC3-1}) (up to a redefinition $z\to -z q^{1/2}$ and identifying $q=q_s$) in a chamber labeled by $k$.


\subsubsection{Conifold}

Here we illustrate a relation between matrix models and open BPS generating functions for the conifold, related to a brane associated to the external leg of a toric diagram, as in fig. \ref{fig-conibrane}. With appropriate choice of framing its amplitude reads  
$$
\cZ_{top}^{open} = \frac{L(z,q_s)}{L(zQ,q_s)}.    
$$
This also reduces to the modular generating function in the non-commutative chamber $n=k=0$. More generally, let us consider open BPS counting corresponding to the closed chamber labeled by $n$, and open chamber labeled by $k$. From the condition (\ref{Zopen}), after the shift  $z\to -z q^{1/2}$, we get a general generating function of open BPS states
\be
\cZ^{open, \, k}_{n} = |\cZ^{open}_{top}|^2_{\ chamber} = 
\prod_{l=1}^{\infty} \frac{(1 + z q_s^l)(1 + z^{-1} q_s^{k+l-1})}{(1 + zQ q_s^l)(1 + z^{-1}Q^{-1} q_s^{n+k+l-1})}. \label{Znk-coni}
\ee

\begin{figure}[htb]
\begin{center}
\includegraphics[width=0.7\textwidth]{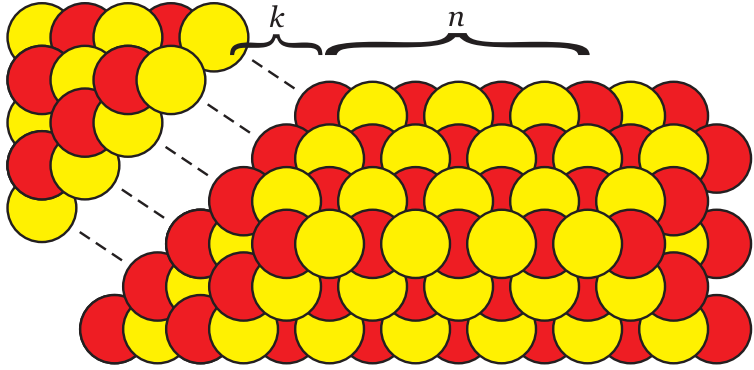} 
\begin{quote}
\caption{\emph{Factorization of the conifold pyramid which leads to open BPS generating functions. The size of the pyramid $n$ represents the closed BPS chamber, while the size $k$ encodes the open BPS chamber.}} \label{fig-coni-split}
\end{quote}
\end{center}
\end{figure}

This result arises from matrix model viewpoint as follows. We take advantage of the results of section \ref{ssec-ExampleConifold}, where we determined $\overline{A}_{\pm}(x) = \G_{\pm}(x) \widehat{Q}_1 \G'_{\pm}(x) \widehat{Q}_0$ and $\overline{W} = \G_{-}(1) \widehat{Q}_1 \G'_{+}(1) \widehat{Q}_0$. 
    This leads to
\be
Z_n  =  \langle 0 | \prod_{i=k}^{\infty}  \overline{A}_+(1) | \mathbb{I} | \prod_{j=0}^{k-1}  \overline{A}_+(1) |\overline{W}^n | \Omega_-\rangle = f^{k}_{n} Z_{matrix}.
\ee
In terms of $\mu=-\frac{1}{q_1}=Q^{-1} q^n$ the matrix integral takes form
$$
Z_{matrix} =  \int \mathcal{D}U \prod_{\alpha} \prod_{j=1}^{\infty} \frac{(1 + z_{\alpha}q^{j})(1 + z_{\alpha}^{-1} q^{k+j-1}) }{(1 + z_{\alpha} \mu q^{j})
(1+  z^{-1}_{\alpha}\mu^{-1} q^{j+n+k-1})}.
$$
The integrand of this matrix model indeed agrees with (\ref{Znk-coni}) M-theory (again identifying $\mu$ with K{\"a}hler parameter  used in M-theory derivation, and setting $q=q_s$). In the limit $n\to\infty$ followed by $\mu\to 0$ we get the result for $\mathbb{C}^3$ given in (\ref{C3open-matrix}). On the other hand, for both $n,k\to\infty$, the integrand reduces to the open topological string amplitude given by a ratio of two quantum dilogarithms. 
The prefactor above is found as
$$
f^{k}_{n} = M(q)^2 \frac{M(\mu q^{k},q) M(Q q^{k},q) }{M(\mu,q) M(q^{k},q) M(Q,q) M(\mu Q q^{k},q)} \prod_{j=1}^{\infty}\big(1 - \mu q^j \big)^n.
$$


\subsubsection{$\C^3/\Z_2$}

As another example we consider open BPS counting functions for resolved $\mathbb{C}^3/\mathbb{Z}_2$ singularity. In this case the topological string partition function for a brane on the external leg is
$$
Z^{ext}_{top} = L(z,q_s) L(zQ,q_s).
$$
This implies the following BPS generating functions in a closed chamber $n$ and open chamber $k$ (after  $z\to -z q^{1/2}$ shift)
$$
\cZ^{open,\, k}_{n} = |\cZ^{open}_{top}|^2 _{chamber} = \prod_{l=1}^{\infty} (1+z q_s^l)(1 + zQ q_s^l)(1+z^{-1} q_s^{k+l-1})(1+z^{-1}Q^{-1} q_s^{n+k+l-1}).
$$

On the other hand, using results of section \ref{ssec-OrbiC3}, i.e.  $\overline{A}_{\pm}(x) = \G_{\pm}(x) Q_1 \G_{\pm}(x) Q_0$  and $\overline{W} = \G_{-}(1) \widehat{Q}_1 \G_{+}(1) \widehat{Q}_0$, and redefining $\mu=\frac{1}{q_1}=Q^{-1} q^n$, we obtain
\bea
Z^n_k & = & \langle 0 | \prod_{i=k}^{\infty}  \overline{A}_+(1) | \mathbb{I} | \prod_{j=0}^{k-1}  \overline{A}_+(1) | \overline{W}^n | \Omega_-\rangle = f^k_n Z_{matrix} = \nonumber \\
& = & f^{k}_{n} \int \mathcal{D}U \prod_{\alpha} \prod_{j=1}^{\infty} (1 + z_{\alpha}q^{j})(1 + z_{\alpha}\mu  q^{j})
(1 + \frac{q^{k+j-1}}{z_{\alpha}})(1+ \frac{q^{n+k+j-1}}{z_{\alpha}\mu}).
\eea
The matrix integrand indeed agrees with the M-theory result (when written in terms of the argument $\mu$) above. The prefactor above reads
$$
f^{k}_{n} = M(q)^2 \frac{M(\mu,q) M(Q,q)}{ M(\mu q^{k},q) M(q^{k},q) M(\mu Q q^{k},q) M(Q q^{k},q)} \prod_{j=1}^{\infty}(1-\mu q^j)^{-n}.
$$


\section{Refined crystals and matrix models}  \label{sec-refine}

In the last section we turn our attention to so-called refined BPS amplitudes, and explain how to incorporate the effect of such refinement in the fermionic formalism and matrix models, following \cite{refine}. To start with, we recall that there are various definitions of refinements, which arise in the context of BPS counting or topological string theory. Here we focus on closed BPS states and consider the following characterization. We introduce an additional parameter $y$ on which multiplicities of D6-D2-D0 states $\Omega$ in the original definition of the generating function (\ref{BPSpartition}) may depend
$$
Z^{ref}_n(q_s,Q) = \sum_{\alpha,\gamma} \Omega^{ref}_{\alpha,\gamma}(n;y) q_s^{\alpha} Q^{\gamma}.
$$
For fixed D0-brane and D2-brane charges $\alpha$ and $\gamma$, and a choice of closed BPS chamber $n$, refined degeneracies are defined as 
\be
\Omega^{ref}_{\alpha,\gamma}(n;y) = \textrm{Tr}_{\mathcal{H}_{\alpha,\gamma}(n)} (-y)^{2J_3},     \label{OmegaTr}
\ee
where $\mathcal{H}_{\alpha,\gamma}(n)$ denotes a space of BPS states with given charges $\alpha,\gamma$ and asymptotic values of moduli corresponding to a chamber $n$, and $J_3$ represents a generator of the spatial rotation group. For $y=1$ these degeneracies reduce to those in (\ref{BPSpartition}). These refined degeneracies are interesting invariants if the underlying Calabi-Yau space does not posses complex structure deformations -- and this is indeed the case for non-compact, toric manifolds we are interested in. In \cite{RefMotQ} it was argued that these invariants agree with motivic Donaldson-Thomas invariants of \cite{KS}, and in the case of the resolved conifold the corresponding BPS generating functions were derived using the refined wall-crossing formula, and encoded in a refined crystal model. From mathematical viewpoint this setup was generalized to the whole class of toric manifolds without compact four-cycles in \cite{Nagao-open}, and shown therein to agree, in the commutative chamber, with refined topological vertex computations. The refined topological vertex was introduced in \cite{refined-vertex}, see also \cite{Taki,refined-Kanno,AwataKanno-refined}. For other formulations of refinement see \cite{Nek,ref-Antoniadis,Yu-refine}. 

Our aim in this section is to construct refined crystal and matrix models, which would encode refined BPS generating functions, and in particular (in the commutative chamber) refined topological string amplitudes. We note that an additional motivation to find such models arises from the AGT conjecture \cite{AGT} and the results of \cite{DV2009}, which state that partition functions of Seiberg-Witten theories in the $\Omega$-background (which are related to topological strings by geometric engineering) are reproduced by matrix models with $\beta$-deformed measure (i.e. with Vandermonde determinant raised to the power $\beta$). Explicit construction of one class of such $\beta$-deformed models, however only to the leading order, was given in \cite{2009betaMatrix}; some other works analyzing five-dimensional beta-deformed models include \cite{MMM-Selberg,AY-beta}. On the other hand, explicit computations for simpler $\beta$-deformed model with gaussian potential \cite{Klemm-Omega,Marino-beta}, revealed that it does not reproduce refined topological string amplitude for the conifold (even though the unrefined topological string partition function is properly reproduced when $\beta=1$, see \cite{Marino,lens-matrix}). Nonetheless, the question whether there exist matrix models which reproduce such refined amplitudes remained valid. As we show below (following \cite{refine}) such models can indeed be constructed by appropriate deformation of the matrix model potential (rather than the measure). We note that recently another class of matrix models (with different than above deformation of the measure) was proposed \cite{mina2011}, which also reproduce refined generating functions.

Let us also note that in this section we consider the same set of walls as in the unrefined case. More general walls, along which only refined BPS states decay, may also exist \cite{RefMotQ}. They are called \emph{invisible walls} and they do not arise in our analysis.

In this section we use the following refined notation. The string coupling $g_s$, related to the D0-brane charge as $q_s = e^{-g_s}$ in the unrefined case, is replaced by two parameters
$$
\epsilon_1 = \sqrt{\beta} g_s, \quad \epsilon_2 = -\frac{g_s}{\sqrt{\beta}}, 
$$
or equivalently $\beta = -\frac{\epsilon_1}{\epsilon_2},  \epsilon_1 \epsilon_2 = -g_s^2$. We also often use the exponentiated parameters
\be 
t_1 = e^{-\epsilon_1}, \quad t_2 = e^{\epsilon_2},    \label{t1t2}
\ee
and introduce 
$$
g_s B = \epsilon_1 + \epsilon_2 = g_s\big(\sqrt{\beta} - \frac{1}{\sqrt{\beta}} \big).
$$
The variable $y$ in (\ref{OmegaTr}) is related to $t_1$ and $t_2$ as  $y=t_1/q_s = q_s/t_2$, so that $y^2=t_1/t_2 = q_s^B$. In this notation the unrefined situation $y=1$ corresponds to $\beta=1$, for which $\epsilon_1=-\epsilon_2=g_s$ and $t_1=t_2=q_s$ and $B=0$.

Let us present now refined BPS generating functions for some Calabi-Yau spaces:

\begin{itemize}
\item
For $\mathbb{C}^3$ we get the refined MacMahon function \cite{refined-vertex}
\be
Z^{\mathbb{C}^3} = M_{ref}(t_1,t_2) = \prod_{k,l=0}^{\infty} \frac{1}{1 - t_1^{k+1} t_2^l}.   \label{Zbps-C3}
\ee
In this case there is no K{\"a}hler parameter, and therefore there are no interesting wall-crossing phenomena.\footnote{In fact one can consider more general family of refinements parametrized by $\delta$, such that $M_{\delta}(t_1,t_2) = \prod_{k,l=0}^{\infty} \big(1 - t_1^{k+1+\frac{\delta-1}{2}} t_2^{l-\frac{\delta-1}{2}}\big)^{-1}$. In this paper we fix the value $\delta=1$ (note that in \cite{RefMotQ} another choice $\delta=0$ was made).} 

\item 
For the resolved conifold refined generating functions were computed in \cite{RefMotQ} using refined wall-crossing formulas. In the closed BPS chamber labeled by $n-1$ these generating functions read
\be
Z^{conifold}_{n-1} = M_{ref}(t_1,t_2)^2 \Big( \prod_{k,l=0}^{\infty} \big(1 - Q t_1^{k+1} t_2^l\big) \Big) \Big( \prod_{k\geq 1,\ l\geq 0, \ k+l\geq n} \big(1 - Q^{-1} t_1^{k} t_2^l\big) \Big) \label{Zbps-coni}.
\ee
In the commutative chamber $n\to\infty$ the terms in the last bracket decouple and the BPS generating function agrees (up to the refined MacMahon factor) with the refined topological string amplitude 
$$
Z^{conifold}_{\infty} = M_{ref}(t_1,t_2) \, \mathcal{Z}^{ref}_{top} = M_{ref}(t_1,t_2) \prod_{k,l=0}^{\infty} (1 - Q t_1^{k+1} t_2^{l})
$$
On the other hand, in the non-commutative chamber $n=0$ the refined generating function is given by the modulus square of the refined topological string amplitude.

\item
For a resolution of $\mathbb{C}^3/\mathbb{Z}_2$ singularity there is also a discrete set of chambers parametrized by an integer $n$. The corresponding BPS generating functions read
\be
Z^{\mathbb{C}^3/\mathbb{Z}_2}_{n-1} = M_{ref}(t_1,t_2)^2 \Big( \prod_{k,l=0}^{\infty} \big(1 - Q t_1^{k+1} t_2^l\big)^{-1} \Big) \Big( \prod_{k\geq 1,\ l\geq 0, \ k+l\geq n} \big(1 - Q^{-1} t_1^{k} t_2^l\big)^{-1} \Big) .  \label{Zbps-C3Z2}
\ee

\item Generating functions for an arbitrary toric geometry, in for the non-commutative chamber, are given (as in the unrefined case) by the modulus square of the (instanton part of the) refined topological string amplitude
\be
Z^{ref}_0 = |\cZ^{ref}_{top}|^2 \equiv \cZ^{ref}_{top}(Q_i) \cZ^{ref}_{top}(Q_i^{-1}).     \label{Z-Ztop2}
\ee
The (instanton part of the) refined topological string amplitude is given by \cite{refined-vertex,Taki}
\be
\cZ^{ref}_{top}(Q_i) =  M_{ref}(t_1,t_2)^{\frac{N+1}{2}} \prod_{k,l=0}^{\infty} \prod_{1\leq i < j \leq N+1} \Big(1- (Q_i Q_{i+1}\cdots Q_{j-1}) \, t_1^{k+1} t_2^l\Big)^{-\tau_i \tau_j}. \label{Ztop-strip-ref}
\ee

\end{itemize}


\subsection{Refining free fermion representation}

\begin{figure}[htb]
\begin{center}
\includegraphics[width=0.4\textwidth]{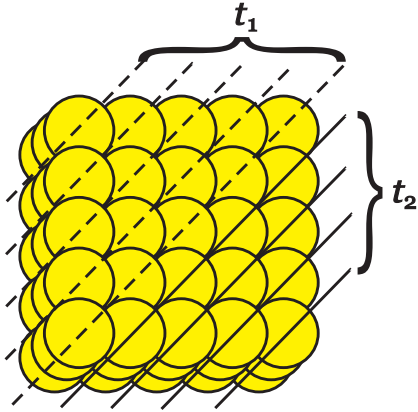} 
\begin{quote}
\caption{\emph{Refined plane partitions which count D6-D0 bound states in $\C^3$. In each slice balls, which intersect a dashed or solid line, have respectively weight $t_1$ or $t_2$. The resulting generating function is the refined MacMahon function $M_{ref}(t_1,t_2)$. }} \label{fig-C3ref}
\end{quote}
\end{center}
\end{figure}

In the non-refined case to a geometry consisting of $N$ $\mathbb{P}^1$'s we associated in section \ref{sec-BPSfermions} a crystal which can be sliced into layers in $N+1$ colors, denoted $q_0, q_1, q_2, \ldots, q_N$. In that case parameters $q_1,\ldots, q_N$ encode K{\"a}hler parameters of the geometry $Q_1,\ldots,Q_N$, while the product $\prod_{i=0}^N q_i$ is mapped to (possibly inverse of) $q_s=e^{-g_s}$. In the refined case the assignment of colors must take into account a refinement of a single parameter $q_s$ into $t_1$ and $t_2$ introduced in (\ref{t1t2}). In particular, in the non-commutative chamber $q_{i\neq 0}$ are mapped (up to a sign, as in the non-refined case) to $Q_i$, however we will have to replace $q_0$ by two refined colors $q_0^{(1)}$ or $q_0^{(2)}$, so that $t_i = q_0^{(i)} q_1\cdots q_N$, for $i=1,2$. The simplest case of $\C^3$ refined plane partitions (discussed also in \cite{refined-vertex}) is shown in fig. \ref{fig-C3ref}. In what follows we will discuss assignment of colors for other manifolds. 

Now we wish to follow the idea of section \ref{sec-BPSfermions}. Firstly, we wish to construct refined states $|\Omega^{ref}_{\pm}\rangle$ whose correlators would reproduce refined BPS amplitudes in the non-commutative chamber
\be
Z^{ref}_0 = \langle \Omega^{ref}_+ | \Omega^{ref}_- \rangle.          \label{Z0ref}
\ee
Secondly, we wish to construct refined wall-crossing operators $\overline{W}^{ref}_n$, such that the BPS generating function in $n$'th chamber can be written as
\be
Z^{ref}_n = \langle \Omega^{ref}_+ | \overline{W}^{ref}_n |\Omega^{ref}_- \rangle.    \label{Znref}
\ee

In section \ref{ssec-ncDT} below we construct states $| \Omega^{ref}_{\pm} \rangle$ for arbitrary manifold in a class of our interest. In section \ref{ssec-coni} we construct states $| \Omega^{ref}_{\pm} \rangle$ and wall-crossing operators $\overline{W}^{ref}_n$ for all chambers of the resolved conifold and a resolution of $\C^3/\Z_2$ singularity.


\subsubsection{Arbitrary geometry -- non-commutative chamber}   \label{ssec-ncDT}

Here we construct fermionic states $| \Omega^{ref}_{\pm} \rangle$, which allow to write the BPS generating functions in the non-commutative chamber as in (\ref{Z0ref}). As in the non-refined case, the states $|\Omega^{ref}_{\pm}\rangle$ are constructed from an interlacing series of vertex operators $\G^{\tau_i}_{\pm}$ and weight operators. The refinement does not modify the three-dimensional shape of the corresponding crystal, therefore the assignment of vertex operators is the same as in the non-refined case (\ref{ti-GammaPM}), as explained in section \ref{sssec-toric}. However, this is assignment of colors, encoded in the weight operators, which is modified in the refined case.  Let us introduce $N$ operators $\widehat{Q}_i$ representing colors $q_i$, for $i=1,\ldots,N$, and in addition two other colors $q_0^{(1)}$ and $q_0^{(2)}$, which are eigenvalues of $\widehat{Q}_0^{(1)}$ and $\widehat{Q}_0^{(2)}$. Operators $\widehat{Q}_1,\ldots,\widehat{Q}_N$, similarly as in section \ref{sssec-toric}, are assigned to $\mathbb{P}^1$'s in the toric diagram, and we introduce
\be 
\widehat{Q}^{(i)} = \widehat{Q}_0^{(i)} \widehat{Q}_1\cdots \widehat{Q}_N,\qquad  t_{i} = q_0^{(i)} q_1 \cdots q_N, \qquad \textrm{for}\ i=1,2.    \label{t12}
\ee
Now we define refined version of $\overline{A}_{\pm}$ operators
\bea
\overline{A}_{+}(x) & = & \G_{+}^{\tau_1} (x) \widehat{Q}_1 \G_{+}^{\tau_2} (x) \widehat{Q}_2 \cdots \G_{+}^{\tau_N} (x) \widehat{Q}_N \G_{+}^{\tau_{N+1}} (x) \widehat{Q}_0^{(1)}   ,       \nonumber        \\
\overline{A}_{-}(x) & = & \G_{-}^{\tau_1} (x) \widehat{Q}_1 \G_{-}^{\tau_2} (x) \widehat{Q}_2 \cdots \G_{-}^{\tau_N} (x) \widehat{Q}_N \G_{-}^{\tau_{N+1}} (x) \widehat{Q}_0^{(2)}.      \nonumber
\eea
Commuting all $\widehat{Q}_i$'s to the left or right it is convenient to use 
\bea
A_+(x) & = & \big(\widehat{Q}^{(1)}\big)^{-1} \, \overline{A}_{+}(x) = \G_{+}^{\tau_1} \big(xt_1\big)  \G_{+}^{\tau_2} \big(\frac{xt_1}{q_1}\big) \G_{+}^{\tau_3} \big(\frac{xt_1}{q_1 q_2}\big) \cdots \G_{+}^{\tau_{N+1}} \big(\frac{xt_1}{q_1q_2\cdots q_N}\big), \label{Aplus-ref} \nonumber  \\
A_-(x) & = & \overline{A}_{-}(x) \, \big(\widehat{Q}^{(2)}\big)^{-1} = \G_{-}^{\tau_1} (x)  \G_{-}^{\tau_2} (xq_1) \G_{-}^{\tau_3} (x q_1 q_2) \cdots \G_{-}^{\tau_{N+1}} (x q_1q_2 \ldots q_N), \label{Aminus-ref}   \nonumber
\eea
and when the argument of these operators is $x=1$ we often use a simplified notation
$$
\overline{A}_{\pm} \equiv \overline{A}_{\pm}(1), \qquad A_{\pm} \equiv A_{\pm}(1). 
$$
Finally we associate to a given toric manifold two (refined) states
\bea
\langle \Omega^{ref}_+| & = & \langle 0 | \ldots \overline{A}_+(1) \overline{A}_+(1) \overline{A}_+(1) = \langle 0 | \ldots A_+(t_1^2) A_+(t_1) A_+(1),  \label{Omega-plus-ref}  \nonumber  \\
| \Omega^{ref}_- \rangle & = & \overline{A}_-(1) \overline{A}_-(1) \overline{A}_-(1) \ldots |0\rangle = A_-(1) A_-(t_2) A_-(t_2^2) \ldots |0\rangle ,  \label{Omega-minus-ref}    \nonumber
\eea
where $|0\rangle$ is the fermionic Fock vacuum.

Our claim now is that the refined BPS generating function can be written as
\be
Z^{ref}_0 = \langle \Omega^{ref}_+ | \Omega^{ref}_- \rangle  \equiv \cZ_{top}(Q_i) \cZ_{top}(Q_i^{-1}),     \label{Z-cZref}
\ee
with $\cZ_{top}(Q_i)$ given in (\ref{Ztop-strip}), and if one identifies $q_i$ parameters which enter a definition of $|\Omega^{ref}_{\pm}\rangle$ and string parameters $Q_i=e^{-T_i}$ (for $i=1,\ldots,N$) as follows
\be
q_i = (\tau_i \tau_{i+1}) Q_i,   \label{qQref}    \nonumber
\ee
and with refined parameters $t_{1,2}$ identified as in (\ref{t12}).

To prove (\ref{Z-cZref}) for general geometry, we first note that commuting operators $A_+(x)$ with $A_-(y)$ 
$$
A_+(x) A_-(y) = A_-(y) A_+(x)\, C(x,y),
$$
gives rise to a factor 
$$
C(x,y) = \frac{1}{(1 - t_1 xy)^{N+1}} \prod_{1\leq i < j \leq N+1} \Big( \big(1 - (\tau_i \tau_j) x y t_1 (q_i q_{i+1}\ldots q_{j-1}) \big) \big(1 -  \frac{(\tau_i \tau_j) x y t_1}{ q_i q_{i+1}\ldots q_{j-1}} \big)  \Big)^{-\tau_i \tau_j}.
$$
Now we write the states $| \Omega^{ref}_{\pm} \rangle$ in terms of $A_{\pm}$ operators, and commute $\Gamma_{\pm}$ within each pair of $A_+$ and $A_-$ separately
$$
Z^{ref}_0 = \langle \Omega^{ref}_+ | \Omega^{ref}_- \rangle = \langle 0| \Big( \prod_{i=0}^{\infty} A_+(t_1^i) \Big) \Big( \prod_{j=0}^{\infty} A_-(t_2^j) \Big) |0\rangle = \prod_{i,j=0}^{\infty} C(t_1^i,t_2^j).
$$
This last product reproduces modulus square (\ref{Z-cZref}) of the refined topological string partition function (\ref{Ztop-strip-ref}), and therefore proves the claim (\ref{Z0ref}). Moreover, for the special $\beta=1$, we automatically obtain the proof of the analogous statement (\ref{Z-Omega}) in the non-refined case from section \ref{ssec-FockState}. 



\subsubsection{Refined conifold and $\C^3/\Z_2$ }   \label{ssec-coni}

\begin{figure}[htb]
\begin{center}
\includegraphics[width=0.7\textwidth]{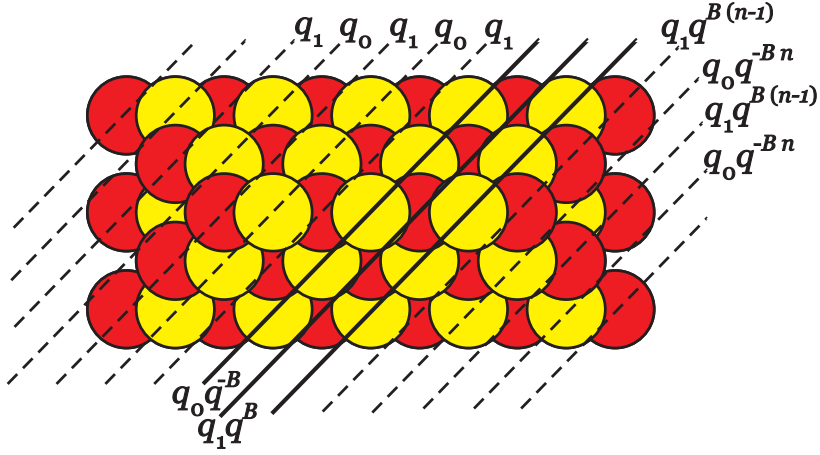} 
\begin{quote}
\caption{\emph{Refined pyramid crystal for the conifold, in the chamber corresponding to $n$ stones in the top row. Along each slice (as indicated by broken or solid lines) all stones have the same color, assigned as follows. On the left side (along broken lines), each light (yellow) and dark (red) slice has color denoted $q_0$ and $q_1$ respectively. Moving to the right, in the intermediate region (along solid lines), a color of each new light or dark slice is modified by respectively $q^{\mp B}$ factor (with respect to the previous light or dark slice). On the right side (again along broken lines), each light or dark slice has again the same color, respectively $q_0 q^{-B n}$ or $q_1 q^{B(n-1)}$. The assignment of colors in the intermediate region (along solid lines) interpolates between constant assignments on the left and right side of the pyramid.  }} \label{fig-colors}
\end{quote}
\end{center}
\end{figure}

\begin{figure}[htb]
\begin{center}
\includegraphics[width=0.4\textwidth]{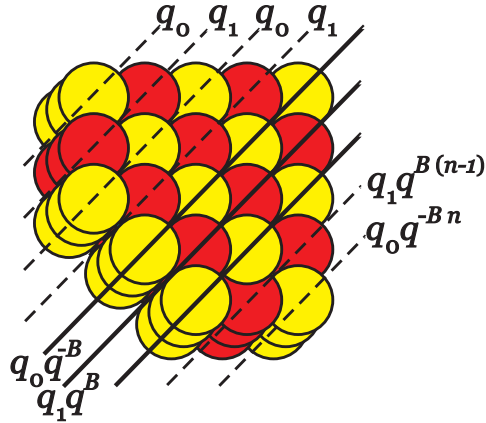} 
\begin{quote}
\caption{\emph{Refined pyramid crystal for the resolution of $\C^3/\Z_2$ singularity, in the chamber corresponding to $n$ stones in the top row, as seen from the bottom (i.e. a negative direction of $z$-axis). Even though the three-dimensional shape of the crystal is different than in the conifold case, the assignment of colors is the same, see fig. \ref{fig-colors}. }} \label{fig-C3Z2}
\end{quote}
\end{center}
\end{figure}

We can now extend the fermionic representation to non-trivial chambers, for simplicity restricting our considerations to the case of a conifold and a resolution of $\mathbb{C}^3/\mathbb{Z}_2$ singularity, which both involve just one K{\"a}hler parameter $Q_1 \equiv Q$. Our task amounts to determining appropriate wall-crossing operators, denoted $\overline{W}^{ref}_{n-1}$, so that in the chamber labeled by $n-1$ the BPS generating function can be written as
\be
Z^{ref}_{n-1} = \langle \Omega^{ref}_+ | \overline{W}^{ref}_{n-1} |\Omega^{ref}_- \rangle.    \label{ZOmegaW}
\ee
In these both cases the toric diagram has two vertices, the first one of type $\tau_1=1$ and the second one denoted now $\tau\equiv \tau_2$, and $\tau=\mp 1$ respectively for the conifold and $\mathbb{C}^3/\mathbb{Z}_2$. A crystal associated to the expression (\ref{ZOmegaW}) has $n$ stones in the top row and can be sliced into interlacing single-colored layers. The assignment of colors is analogous as in the pyramid model discussed in \cite{RefMotQ,Nagao-open}. The pyramid crystal for the conifold and $\C^3/\Z_2$ are shown in fig. \ref{fig-colors} and \ref{fig-C3Z2}. 

The assignment of colors is determined as follows. All stones on one side of the crystal are encoded in  
$$
\langle \Omega^{ref}_+ | = \langle 0| \ldots \Big(\G_+(1) \widehat{Q}_1 \G_+^{\tau}(1) \widehat{Q}_0  \Big) \Big(\G_+(1) \widehat{Q}_1 \G_+^{\tau}(1) \widehat{Q}_0  \Big).
$$ 
The K{\"a}hler parameter $Q$, as well as the parameter $t_1$, arise as
$$
q_1 = \tau Q t_1^{1-n},\qquad q_0 = \tau \frac{t_1^n}{Q},\qquad \textrm{so that} \quad q_0 q_1=t_1.
$$

Now the crystal with $n-1$ additional stones in the top row arises from an insertion of the operator
\bea
\overline{W}^{ref}_{n-1} & = & \Big(\G_-(1) \widehat{Q}_1 \G_+^{\tau}(1) \widehat{Q}_0 \widehat{q^{-B}} \Big) \Big(\G_-(1) \widehat{Q}_1 \widehat{q^{B}} \G_+^{\tau}(1) \widehat{Q}_0 \widehat{q^{-2B}} \Big) \ldots \nonumber \\
& & \ldots \Big(\G_-(1) \widehat{Q}_1 \widehat{q^{(n-2)B}} \G_+^{\tau}(1) \widehat{Q}_0 \widehat{q^{(1-n)B}} \Big).  \nonumber
\eea
This operator is made of $n-1$ terms of the form $\Big(\G_-(1) \widehat{Q}_1 \widehat{q^{i B}} \G_+^{\tau}(1) \widehat{Q}_0 \widehat{q^{-(i+1)B}} \Big)$ for $i=0,\ldots,n-2$, where in each subsequent dark or light slice we insert one additional operator $\widehat{q^{\pm B}}$. This additional operator changes the weight of each stone in this slice by $q^{\pm B} = (t_1/t_2)^{\pm 1}$ (with respect to the previous slice of the same light or dark color).

Finally, all stones on the right side of the crystal have again the same light or dark color, so that the corresponding state is
$$
|\Omega^{ref}_-\rangle =  \Big(\G_-(1) \widehat{Q}_1 \widehat{q^{(n-1)B}} \G_-^{\tau}(1) \widehat{Q}_0 \widehat{q^{-n B}} \Big) \Big(\G_-(1) \widehat{Q}_1 \widehat{q^{(n-1)B}} \G_-^{\tau}(1) \widehat{Q}_0 \widehat{q^{-n B}} \Big) \ldots |0\rangle.
$$
We see that varying weights in the middle range (along solid lines in fig. \ref{fig-colors} and \ref{fig-C3Z2}) interpolate between fixed weights of light and dark stones on two external sides of a crystal.

We can now commute away all weight operators in the above expressions, using commutation relations from section \ref{ssec-fermion}. This results in 
\be
Z^{ref}_{n-1} = \langle 0| \Big( \prod_{k=1}^{\infty} \G_+(t_1^k) \G_+^{\tau} (t_1^k/q_1) \Big)  \Big(\prod_{i=0}^{n-2} \G_-(t_2^i) \G_+^{\tau} (q_1^{-1} t_1^{-i}) \Big) \Big( \prod_{k=0}^{\infty} \G_-(t_2^{n-1+k}) \G_-^{\tau} (tQ t_2^k) \Big) |0\rangle  .   \label{Zconi-ref}
\ee
To check that this is a correct representation we commute all vertex operators, and find
\be
Z^{ref}_{n-1} = M_{ref}(t_1,t_2)^2 \, \prod_{k=1,l=0}^{\infty} (1 - Q t_1^{k} t_2^{l})^{-\tau} \, \prod_{k\geq 1,l\geq 0, k+l\geq n}^{\infty} (1 - Q^{-1} t_1^{k} t_2^{l})^{-\tau},
\ee
where $\tau=\mp 1$ respectively for the conifold and $\mathbb{C}^3/\mathbb{Z}_2$. This result reproduces (\ref{Zbps-coni}) and (\ref{Zbps-C3Z2}), which confirms that the fermionic representation we started with is correct.


\subsection{Refined matrix models}

In the refined case one can associate matrix models to refined generating functions in the same way as described in section \ref{ssec-matrixFermion}, i.e. by inserting the representation (\ref{identity}) of the identity operator into fermionic representation 
(\ref{Z-cZref}) or (\ref{ZOmegaW}). This does not change a unitary character of the matrix model, which is a consequence of the representation (\ref{identity}). However, due to more subtle weight assignments, this is matrix potential which gets deformed by $\beta$-dependent factors. In general we will therefore obtain matrix models of the following form
\be
Z^{ref}_n = f_n\,\int \mathcal{D}U \prod_k e^{-\frac{\sqrt{\beta}}{g_s} V(z_k;\beta)},   \label{Zmatrix-ref}
\ee
where for convenience we introduced a factor $\sqrt{\beta}$ in front of the potential $V(z;\beta)$. We will consider a few examples below.

\subsubsection{Non-commutative chamber}

For arbitrary geometry, in the non-commutative chamber, refined matrix model integrand can be expressed in terms of the refined theta function
\be
\Theta(z;t_1,t_2) = \prod_{j=0}^{\infty} (1 + z t_1^{j+1})(1+t_2^{j}/z).  \nonumber
\ee
Repeating the computation described in section \ref{ssec-matrixFermion}, however starting with the refined representation  (\ref{Z-cZref}), in the non-commutative chamber for general geometry we find the matrix model
\be
Z^{ref}_0 = \int \mathcal{D}U \prod_k \prod_{l=0}^N \Theta\big(\frac{\tau_{l+1} z_k }{q_1\cdots q_l} ; t_1,t_2\big)^{\tau_{l+1}} ,  \label{Znoncom-ref}
\ee
i.e. we identify $e^{-\frac{\sqrt{\beta}}{g_s} V(z;\beta)} \equiv \prod_{l=0}^N \Theta\big(\tau_{l+1} z (q_1\cdots q_l)^{-1} ; t_1,t_2 \big)^{\tau_{l+1}}$. The product over $l$ runs over all vertices and we identify K{\"a}hler parameters $Q_p$ with weights $q_p$ via $q_p=(\tau_p \tau_{p+1})Q_p$.


\subsubsection{Refined $\C^3$ matrix model}

We obtain a refined matrix model for $\C^3$ as the special case of (\ref{Znoncom-ref}). For the refined $\mathbb{C}^3$ the BPS generating function is a refined MacMahon function $Z^{ref}=M_{ref}(t_1,t_2)$ introduced in (\ref{Zbps-C3}), and the corresponding matrix integrand takes form of a a refined theta function
\be
e^{-\frac{\sqrt{\beta}}{g_s} V(z;\beta)}  =  \prod_{j=0}^{\infty} (1 + zt_1^{j+1})(1 + t_2^{j}/z) = \Theta(z;t_1,t_2)  . \label{matrixC3}
\ee
Using the asymptotics (\ref{qdilog-expand})
we find the leading order expansion of the potential
\be
e^{-\frac{\sqrt{\beta}}{g_s} V(z;\beta)}  = e^{-\frac{\sqrt{\beta}}{g_s} \big[ -\frac{1}{2}(\log z)^2 - (1-\beta^{-1})\textrm{Li}_2(-z)   + \mathcal{O}(g_s,\beta)  \big] }.    \label{C3-V0}
\ee
The quadratic term in the potential is the same as in the non-refined case. The term involving Li$_2(-z)$, as well as all higher order terms $\mathcal{O}(g_s,\beta)$, arise as deformations which vanish for $\beta=1$. Therefore, for $\beta=1$, we obtain a Chern-Simons matrix model which indeed gives rise to MacMahon function in $N\to\infty$ limit, as we explained in section \ref{ssec-matrix-conifold}. For arbitrary $\beta$, the resolvent $\omega(p)$ can also be found using the Migdal integral (\ref{Migdal}). In principle one could repeat the computation described in section \ref{ssec-matrix-conifold}, however this is technically more involved. Nonetheless, this would lead to $\beta$-deformed end-points of the cut (\ref{aAll}) and (\ref{bAll}), and in consequence to the $\beta$-deformed spectral curve. This curve would be some $\beta$-deformation of the mirror curve given in (\ref{mirrorcurves}). It is still an interesting question to find this curve in the exact form and analyze its properties.


\subsubsection{Refined conifold matrix model}

Finally we find matrix models for the refined conifold. Starting with the representation (\ref{Zconi-ref}), inserting the identity representation (\ref{identity}) and following standard by now computations, we find the following matrix model for the conifold in the $n$'th chamber (corresponding to a pyramid with $(n+1)$ stones on top)
\bea
Z^{ref}_{n} & = & M_{ref}(t_1,t_2)^2 \, \prod_{k=1,l=0}^{\infty} (1 - Q t_1^{k} t_2^{l}) \, \prod_{k\geq 1,l\geq 0, k+l\geq n+1}^{\infty} (1 - Q^{-1} t_1^{k} t_2^{l}) = \label{MM-Zn-conifold-ref}    \nonumber  \\
& = & f_{n}(q,Q) \int \mathcal{D}U \prod_k \prod_{j=0}^{\infty}  \frac{ (1+z_k t_1^{j+1})\ (1+t_2^{j}/z_k)}{ (1+z_k t_1^{j+n+1}/Q)\ (1+t_2^j Q/z_k)},    \nonumber
\eea
with the prefactor given by
\be
f_{n}(q,Q) = \Big( \prod_{i=1}^{n} \prod_{k=0}^{\infty} \frac{1}{1-t_1^{i} t_2^k} \Big) \Big( \prod_{i=1}^{n} \prod_{j=n+1-i}^{\infty} (1-t_1^{i} t_2^j/Q) \Big).    \nonumber    \label{fn-conifold-ref}
\ee
In the limit of the commutative chamber, $n\to \infty$, we get $f_{\infty} = M_{ref}(t_1,t_2)$. Therefore in the commutative chamber we get a matrix model representation of the refined topological string conifold amplitude
\bea
\mathcal{Z}^{ref}_{top} & = & M_{ref}(t_1,t_2) \prod_{k,l=0}^{\infty} (1 - Q t_1^{k+1} t_2^{l}) = \nonumber \\
& = & \int \mathcal{D}U \prod_k  \prod_{j=0}^{\infty}  \frac{ (1+z_k t_1^{j+1})\ (1+t_2^{j}/z_k)}{ (1+t_2^j Q/z_k)}.    \nonumber
\eea
In this case the lowest order potential is a modification of the $\C^3$ potential (\ref{C3-V0}) by a $Q$-dependent dilogarithm term
\be
V(z;\beta) = -\frac{1}{2}(\log z)^2 - (1-\beta^{-1})\textrm{Li}_2(-z) - \textrm{Li}_2(-Q/z) + \mathcal{O}(g_s,\beta).   \label{VconiTop}
\ee
This is quite an interesting result -- as we already explained above, it has been postulated for some time that the refined topological string amplitude for the conifold should have matrix model representation, however it was not clear how to derive it. Here we find an explicit matrix model representation of this amplitude. The corresponding spectral curve would again be a $\beta$-deformation of the conifold mirror curve (\ref{mirrorcurves}). It would be interesting to compare it with other notions of deformed, or quantum mirror curves in literature. We also note that in the limit $Q\to 0$ the above topological string partition function becomes just the refined MacMahon function, and the matrix integral consistently reproduces  $\C^3$ result (\ref{matrixC3}).



\newpage
\begin{center}
{\bf Acknowledgments}
\end{center}

\medskip

I thank Robbert Dijkgraaf, Hirosi Ooguri, Cumrun Vafa and Masahito Yamazaki for discussions and collaboration on related projects. This research was supported by the DOE grant DE-FG03-92ER40701FG-02 and the European Commission under the Marie-Curie International Outgoing Fellowship Programme. The contents of this publication reflect only the views of the author and not the views of the funding agencies.


\end{document}